\documentclass[draft]{agujournal2019}
\usepackage{url} 
\usepackage{lineno}
\usepackage[inline]{trackchanges} 
\usepackage{soul}
\usepackage{amssymb}
\usepackage{amsfonts}
\usepackage{multirow}
\drafttrue

\journalname{Earth and Space Science}

\begin{document}

%
%


\title{Feasibility of Passive Sounding of Uranian Moons using Uranian Kilometric Radiation}

%
%

\authors{A. Romero-Wolf\affil{1}, G. Steinbr\"ugge\affil{1}, J. Castillo-Rogez\affil{1}, C.J. Cochrane\affil{1}, T.A. Nordheim\affil{1}, K.L. Mitchell\affil{1}, N.S. Wolfenbarger\affil{2}, D.M. Schroeder\affil{2}, S. Peters\affil{3}}

\affiliation{1}{Jet Propulsion Laboratory, California Institute of Technology, Pasadena CA USA}
\affiliation{2}{Department of Geophysics, Stanford University, Stanford, CA, USA}
\affiliation{3}{Department of Physics, Naval Postgraduate School, Monterey, CA, USA}

\correspondingauthor{A. Romero-Wolf}{andrew.romero-wolf@jpl.nasa.gov}





\begin{keypoints}
\item Passive Sounding of the Uranian moons is a promising concept and no insurmountable obstacles have been identified.
\item In the presence of NH$_3$ rich oceans, the direct detection of the ice-ocean interface is feasible. 
\item The method works best for the outer moons and low ocean temperatures, making it complementary to magnetic induction measurements.
\end{keypoints}

\justify 
\begin{abstract}
We present a feasibility study for passive sounding of Uranian icy moons using Uranian Kilometric Radio (UKR) emissions in the 100 - 900 kHz band. We provide a summary description of the observation geometry, the UKR characteristics, and estimate the sensitivity for an instrument analogous to the Cassini Radio Plasma Wave Science (RPWS) but with a modified receiver digitizer and signal processing chain. We show that the concept has the potential to directly and unambiguously detect cold oceans within Uranian satellites and provide strong constraints on the interior structure in the presence of warm or no oceans. As part of a geophysical payload, the concept could therefore have a key role in the detection of oceans within the Uranian satellites. The main limitation of the concept is coherence losses attributed to the extended source size of the UKR and dependence on the illumination geometry. These factors represent constraints on the tour design of a future Uranus mission in terms of flyby altitudes and encounter timing. 
\end{abstract}

\section*{Plain Language Summary}
The large moons of Uranus are hypothesized to have subsurface oceans beneath their icy crust. This paper analyzes the possibility to use natural radio emissions originating from Uranian auroras to probe for these oceans. Cold ice is transparent to radio waves allowing reflections from liquid water to be readily observed. Monitoring the radio noise patterns from Uranus and searching for the reflections could constitute a direct way to detect the subsurface oceans.

%
%

\section{Introduction}
The Uranian system consists of the ice giant Uranus and its 27 known moons. Among these moons, the five largest ones Miranda, Ariel, Umbriel, Titania, and Oberon are of particular interest due to their potential for subsurface oceans \cite{Hussmann2006,hendrix2019,bierson2022,Castillo2023}. This possibility is of great interest in the search for potentially habitable environments in the Solar System and could provide insight into the thermal and evolutionary history of the moons. The \textit{Origins, Worlds, and Life} decadal survey prioritized the Uranus Orbiter and Probe mission as the next Flagship to be started this decade. 

Miranda, the innermost of the five moons, is known for its relatively young surface and extensive tectonic features, including cliffs, canyons, and grooves, which have been interpreted as evidence of a recent tidal heating event \cite{beddingfield2015,beddingfield2022miranda}. Ariel also exhibits signs of past activity, most prominently the chasmata, 
canyons likely formed by extension \cite{beddingfield2022ariel}. Umbriel has a cratered surface with little evidence of tectonic activity \cite{schenk2020}. Titania, the second outermost moon, exhibits a mixture of cratered and smooth regions and is less heavily cratered than the surfaces of either Oberon or Umbriel, implying a younger surface \cite{Kirchoff_2022}. Oberon's surface is the most heavily cratered of all the Uranian moons and might therefore have the most ancient surface of the Uranian satellites \cite{Kirchoff_2022}. 

A proven technique for detecting ice-ocean interfaces is magnetic sounding which has been used to discover subsurface liquid water oceans within Europa, Callisto, and Ganymede \cite{Kivelson_1999, Kivelson_2002} as well as a putative magma ocean beneath the volcanically active surface of Io \cite{Khurana_2011}. Magnetic sounding of the Jovian moons is achieved through magnetic induction, which is facilitated by the time varying Jovian magnetic environment in which they are immersed. The two upcoming missions - NASA's Europa Clipper and ESA's JUICE - will further use magnetic sounding to characterize the oceans within Europa, Ganymede, and Callisto \cite{grasset2013,phillips2014}. The strong and highly dynamic magnetic environment of Uranus' magnetosphere also provides a fortuitous laboratory to perform magnetic induction investigation of the Uranian moons. Several recent studies have demonstrated the feasibility to detect induced magnetic field signatures from sub-surface oceans on Uranus' major moons for a wide range of possible ocean configurations~ \cite{Arridge_2021, Cochrane_2021, Weiss_2021}. However, \cite{Castillo2023} showed that sub-surface oceans on these moons, if they exist, could be cold, only a few tens of kilometers thick, and enriched in ammonia. At these conditions, the electrical conductivity of these residual oceans could be very low, which would make them difficult to detect via magnetic induction. 

Passive radar sounding using Uranian Kilometric Radio (UKR) emissions has the potential to provide information about the internal structure of these moons, including the thickness of the ice shell and the presence and depth of subsurface oceans, thus making it a complementary technique to magnetic sounding. Kilometric radio emissions range from 1 to 10 kilometers in wavelength, and are emitted by all planets with substantial atmospheres and magnetic fields \cite{Zarka_2004}.
Kilometric emissions have been observed to originate from Uranus and are hypothesized to be generated by cyclotron maser instability \cite{gulkis1987}. The use of passive radar sounding techniques involves detecting and analyzing the reflection of naturally occurring radio waves off of the surface or subsurface of a geophysical target~\cite{Romero-Wolf_2015}. By analyzing the reflection of radio waves off of the surface or subsurface of Miranda, Ariel, Umbriel, Titania, and Oberon, it may be possible to determine the presence of subsurface oceans and the thickness of the ice shell.

In this paper we will establish the feasibility to passively radar sound  oceans in the subsurface of Uranian moons using the UKR emissions in the 100 - 900 kHz frequency band (wavelengths 0.33-3 km) as a source. This study assumes an instrument with similar specifications as the Cassini Radio and Plasma Wave Science~\cite{Gurnett_2004} but with a modified digitizer and signal processing chain to perform the cross-correlation of the data needed for passive sounding. The study is analogous to the passive sounding feasibility studies done for Jovian moons using Jovian radio bursts~\cite{Romero-Wolf_2015, Schroeder_2016, Steinbruegge_2021}. In Section \ref{sec:concept} we will provide a concept overview and then summarize the current knowledge of the radio source properties. As the sensitivity for passive sounding depends on the spatial extent of the source (which results in coherence losses), the time bandwidth product available to the instrument, the availability of the source, and the radio losses of the medium being probed, Sections~\ref{sec:source}, \ref{sec:receiver}, and \ref{sec:target} present our analysis to address the source properties, the receiver model, and the target properties, respectively to provide an initial validation of the concept. 

\section{Concept Overview}
\label{sec:concept}

The Uranian passive sounder concept is based on prior concepts for sounding of Jupiter's Galilean moons using Jovian radio bursts~\cite{Romero-Wolf_2015, Schroeder_2016, Steinbruegge_2021}. Passive sounding has been demonstrated experimentally on Earth using reflections of the Sun's quiescent radio emissions reflected off the ocean~\cite{Peters_2018}, sand~\cite{Peters_2021a}, and water beneath 1 km of ice in Greenland~\cite{Peters_2021b}. Importantly, the authors demonstrated that synthetic aperture radar (SAR) processing was possible in passive radar sounding, enabling additional gain to be recovered~\cite{Peters_2021a}.

\begin{figure}[ht]
\centering
\noindent\includegraphics[width=0.8\textwidth]{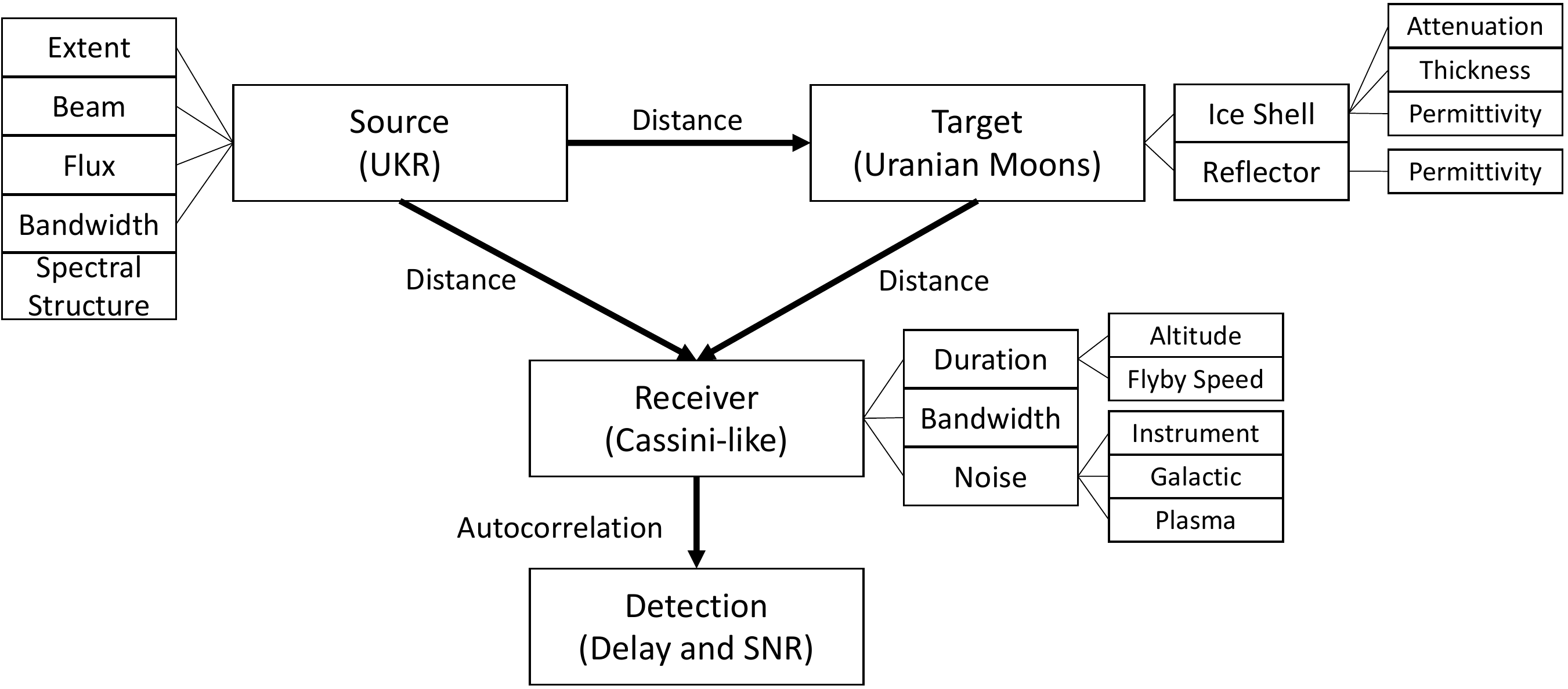}
\caption{Flow chart outlining the passive sounding concept for Uranian icy moons using Uranian kilometric radio (UKR) radio emission. }
\label{fig:concept}
\end{figure}

The passive sounding concept is summarized in Figure~\ref{fig:concept}. The three main components are the UKR source, whose direct emission is recorded by the receiver, and the emission reflected by the Uranian icy moon target which is recorded by the same receiver. 

The source properties relevant for estimating the passive sounding sensitivity are the spatial extent, beam pattern, flux, and instantaneous bandwidth of the radio emissions. These components and their impact on sensitivity will be treated in detail in \S\ref{sec:source}. The spectral structure and its temporal variation can also induce undesired affects to passive sounding~\cite{Carrer_2021}. However, \citeA{Roberts_2022} demonstrated a signal conditioning process that removes the undesired effects of spectral variability by flattening the spectral amplitude modulations at “ripple periods” sufficiently to remove those from the expected echoes. This technique works best in the strong signal regime relevant to this concept and will not be treated further.  

The receiver point-model used for this study is similar to Cassini~\cite{Gurnett_2004} but with a different back-end digitizing at higher instantaneous bandwidth and capable of performing the correlation between the direct and reflected emissions. In the case of a receiver orbiting Uranus, the parameters dominating sensitivity are the duration of the data capture, which is limited by the moon flyby speed and altitude, the receiver's instantaneous bandwidth and center frequency, and the background noise, which we will demonstrate is negligible compared to the UKR itself. The receiver is described in more detail in \S\ref{sec:receiver}.

The key target properties for sensitivity estimates are the moon's ice shell and subsurface reflector properties. The ice shell thickness and attenuation are based on geophysical models for each icy moon of interest (Miranda, Ariel, Umbriel, Titania, and Oberon). The reflected signal strength is also determined by the dielectric contrast between the ice-shell at the interface with the subsurface reflector (e.g. liquid water or bedrock). The icy moon models will be treated in \S\ref{sec:target}.

Other radio frequency measurements that could aid the geophysical interpretation of the data are goniopolarimetric localization~\cite{Cecconi_2009} as performed by Cassini on Saturn, and occultations~\cite{Cecconi_2021}. Goniopolarimetric localization, where the direction of a circularly polarized wave is identified using correlations between co-located antennas with different orientations, allows for the identification of the source position, which is important for estimating the depth of the subsurface reflector. Occultation of the UKR source by the thick ice shells of Uranian icy moons ($\sim100$~km) could potentially be used to estimate the attenuation profiles of the ice. These will be discussed in \S\ref{sec:discussion}.

The models described above will be combined to produce sensitivity estimates and predictions of what the data might look like for a variety of icy moon geophysical scenario point models. The models and predictions are treated in \S\ref{sec:outcomes}. 

\section{Source Properties}
\label{sec:source}

\begin{figure}[ht]
\centering\includegraphics[width=0.6\textwidth]{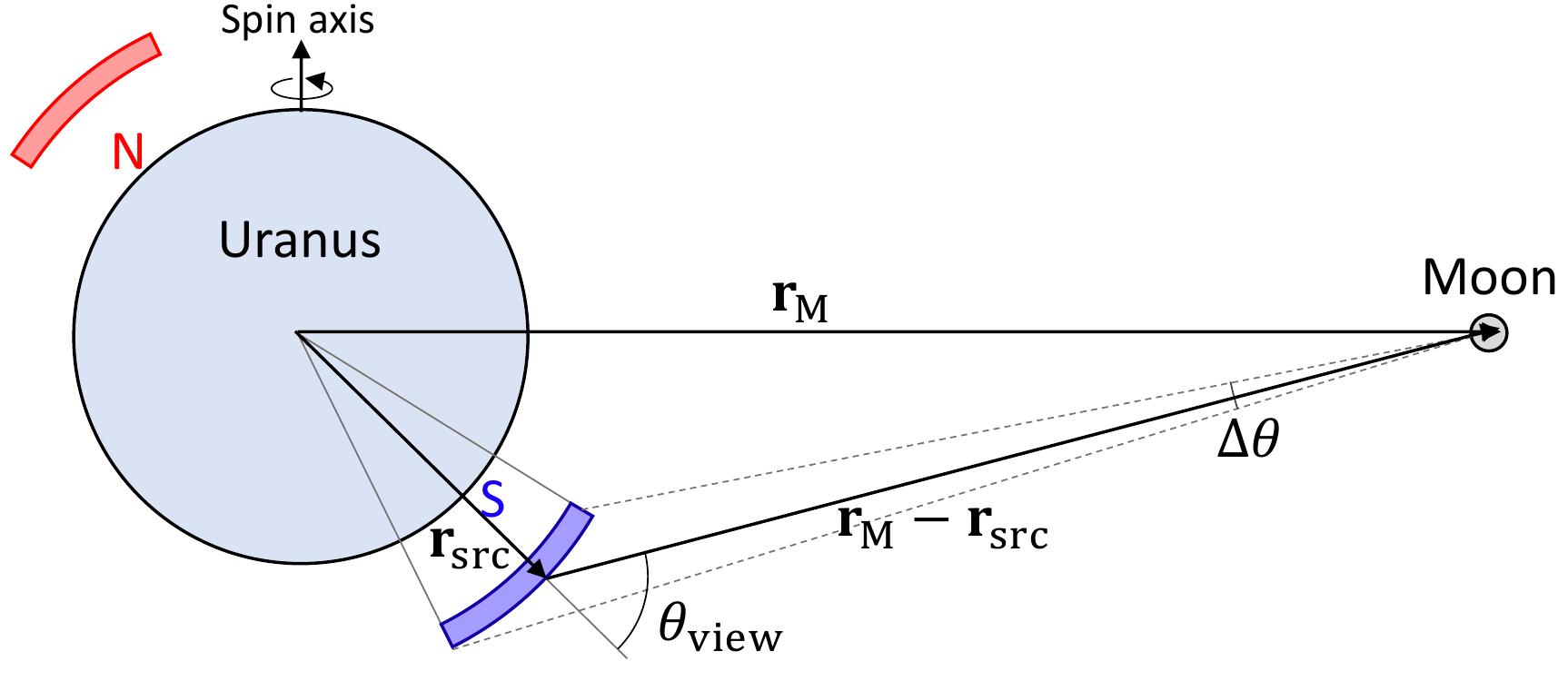}
\caption{Geometry of the UKR sources and icy moon. The location and size of the southern source, labeled by a blue ``S", is based on Voyager-2 observations~\cite{Menietti_1990}. The figure is drawn to scale for Uranus, the source extent and the icy moon Miranda. The northern UKR source, labeled with a red ``N", was not observed by Voyager-2 and we model it as an antipodal clone of the southern source. Ultraviolet images of Uranus taken with the Hubble Space Telescope result in morphological differences between the northern and southern aurora~\cite{Lamy_2017}, which is indicative of differences between their corresponding radio sources. Detailed modeling of the radio source is left to future work (see \S\ref{sec:discussion}) and observational constraints could eventually be obtained directly by a spacecraft. The vector $\mathbf{r}_\mathrm{src}$ corresponds to a location in the UKR source region with the view angle $\theta_\mathrm{view}$ corresponding to the view angle from the icy moon as seen from position $\mathbf{r}_\mathrm{M}$. In this illustration, the icy moon is located at the same longitude as the UKR southern source although this is not necessarily the case. The angle $\Delta\theta$ represents the source extent as seen from the icy moon.
}
\label{fig:geometry}
\end{figure}

The geometric model of the UKR sources with properties relevant for passive sounding is shown in Figure~\ref{fig:geometry}. The UKR sources are located around the northern and southern magnetic poles, which are not aligned with the spin axis of Uranus. The radio emission regions are highly extended with cone-shaped beams emanating along the magnetic field lines. The key properties for passive sounding are the UKR flux (\S\ref{sec:flux}), the angular extent of the source emitting region $\Delta\theta$, which limits the coherence of the correlation used for passive sounding (\S\ref{sec:extent}), and the beam pattern, which limits the view angles $\theta_\mathrm{view}$ for which the source illuminates the icy moon (\S\ref{sec:beam}). 

\subsection{Flux Density}
\label{sec:flux}
\begin{figure}[ht]
\centering
\noindent\includegraphics[width=0.8\textwidth]{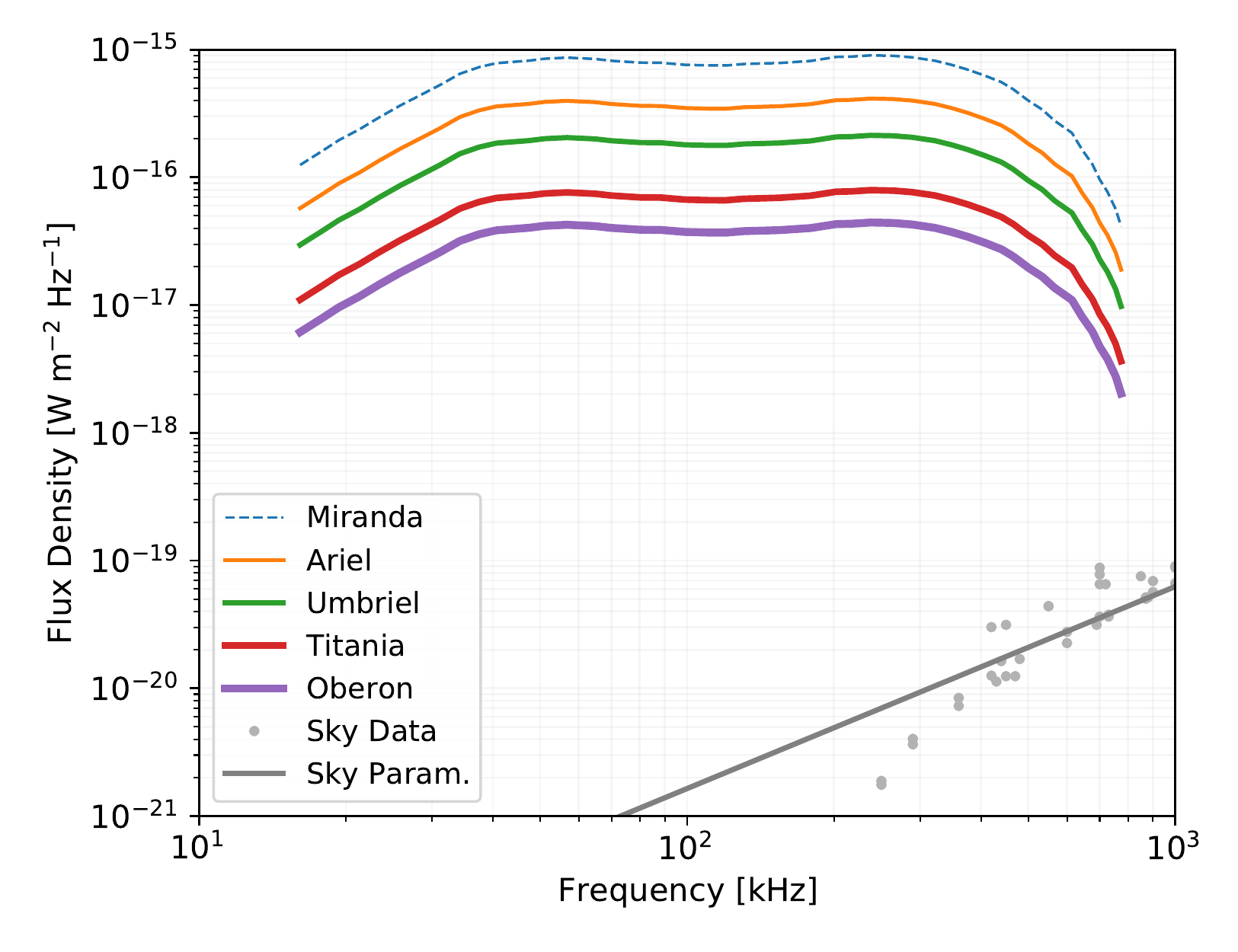}
\caption{The flux density of the Uranian Kilometric Radio (UKR) source~\cite{Zarka_1998} normalized to the distances of the Uranian icy moons. The flux curve for Miranda is dashed because it is currently uncertain whether the beam pattern illuminates it or not. The sky background noise flux (data and parametrization from~\citeA{Cane_1979}) is included for comparison.} 
\label{fig:flux}
\end{figure}

Studies of the UKR source are all based on the encounter by Voyager-2 in January of 1986~\cite{Stone_1987}. In the vicinity of the Uranian icy moons, the UKR is the brightest source in the sky by far in the 25 kHz -- 900 kHz band. In Figure~\ref{fig:flux} we show the average UKR flux from~\citeA{Zarka_1998} normalized to the locations of the Uranian icy moons. Miranda, the icy moon closest to Uranus, is shown in dashed line to indicate that it is uncertain whether the UKR illuminates this moon or not (see \S\ref{sec:beam}). The fluxes incident on the other four moons are several orders of magnitude stronger than the Galactic sky background radiation. In \S\ref{sec:noise} we provide a detailed analysis of background noise sources to show that the limiting background for sounding is the UKR itself. 

\subsection{Extent and Coherence Losses}
\label{sec:extent}
The size of the UKR source is a key parameter to estimate the feasibility of passive sounding. If the spatial extent of an incoherent source is too large, the different emission regions can interfere with each other to the point of removing all coherence in the cross correlation between the directed and reflected radiation, making passive sounding less effective~\cite{Peters_2022}. 

\citeA{Menietti_1990} performed a ray tracing study to determine the southern source region of the smooth high-frequency nightside Uranus kilometric radiation. Their results show that the relevant altitude of the radio source is about 1.5~$R_U$ for 700~kHz frequency. Figure 3 of that paper bounds the spatial extent of the source between 0.47~--~0.53~$R_U$. Here we assume a conservative bound assuming all regions radiate isotropically. We know this is conservative because the radiation follows a conical beam pattern with opening angle spanning from 90$^\circ$ to at least 120$^\circ$ but possibly as large as 160$^\circ$~\cite{Menietti_1990}. Including this more detailed model will improve coherence limitations on the icy moons, except possibly for Miranda since it could reduce its overall illumination. 

\begin{figure}[ht]
\centering
\noindent\includegraphics[width=0.8\textwidth]{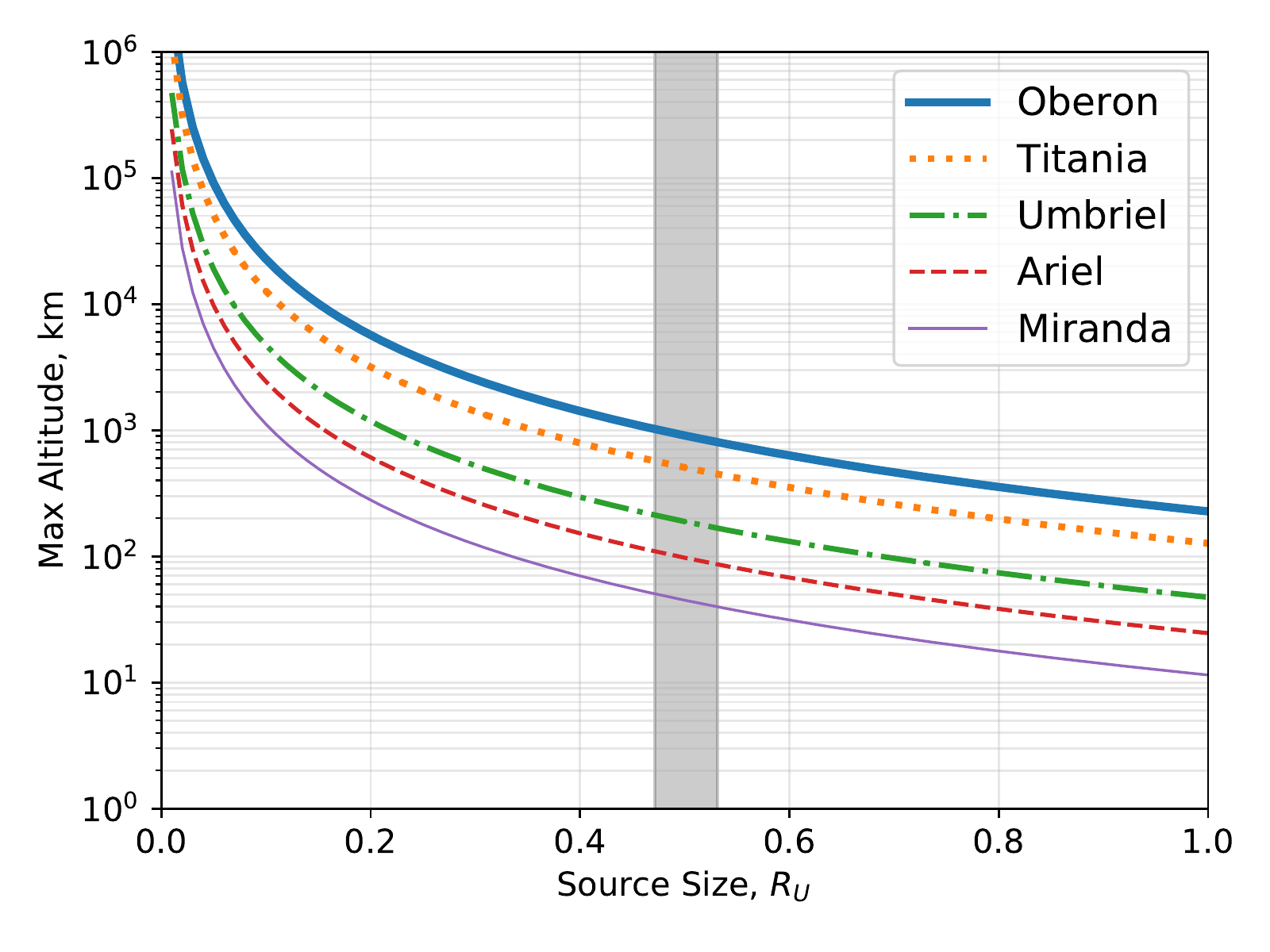}
\caption{Maximum sounding altitude for a reference frequency of 700~kHz. The lines corresponding to each icy moon estimate the maximum altitude at which sounding is viable before coherence losses begin to take place as a function of source size (in units of Uranian radius $R_U$). At 700~kHz, the upper bound on the source size is shown by the gray shaded region. Maximum altitude ranges from $\sim 50$~km for Miranda and are as high as $\sim 1000$~km for Oberon. }
\label{fig:max_altitude}
\end{figure}

The source extent as seen from the observer results in an angular extent of the source denoted by $\Delta\theta$ (see Figure~\ref{fig:geometry}). The estimates below follow~\cite{Peters_2022}.  At a given wavelength $\lambda$, this angular extent determines the maximum altitude $h_\mathrm{max}$ at which a receiver can correlate the direct and reflected signals without significant losses 
\begin{equation}
h_\mathrm{max} =\frac{\lambda}{2(1-\cos\Delta\theta)}.
\end{equation}
The value of $\Delta\theta \simeq \Delta S/D$, where $\Delta S$ is the spatial extent of the source projected in the direction of the icy moon and $D$ is the distance between the icy moon and the UKR source. Figure~\ref{fig:max_altitude} shows estimates of the maximum altitude below which the passive sounding technique will not suffer from coherence losses. The shaded region corresponds to the estimated source size in units of the Uranian radius. The results indicate the fairly low-altitude flybys are required for the closest icy moons ($<$~50~km for Miranda, $<$~110~km for Ariel, $<$~210~km for Umbriel) while higher altitude flybys can be tolerated for Titania and Oberon ($<$~580~km and $<$~1000~km, respectively). 

One limitation of this estimate is that the source is directly overhead. The coherence degrades away from that. Given the source is on for a significant fraction of time, it may be possible to coordinate such a flyby. Note that we also assumed the entire region in Figure~\ref{fig:max_altitude} is contributing to the radiation at any given instance (i.e. the emission at each point is isotropic).  
This is an overestimation since the sources are extended but beamed, which results in a smaller effective source size. More detailed estimates including these effects will be the subject of future work. 

\subsection{Beam Pattern and Target Illumination}
\label{sec:beam}
The beam pattern of the UKR determines the spatio-temporal illumination characteristics of the icy moons. The analysis of~\citeA{Menietti_1990} shows the southern source extends from 30$^\circ$ - 60$^\circ$ in latitude and has a hollow cone beam pattern with opening angle spanning from 90$^\circ$ to at least 120$^\circ$ but possibly as large as 160$^\circ$. The range of view angles $\theta_\mathrm{view}$ (see Figure~\ref{fig:geometry}) corresponding to beam pattern illumination range from 45$^\circ$ (corresponding to the 90$^\circ$ cone opening angle) up to at least 60$^\circ$ and possibly as high as 80$^\circ$.



\begin{figure}[ht]
\noindent\includegraphics[width=0.49\textwidth]{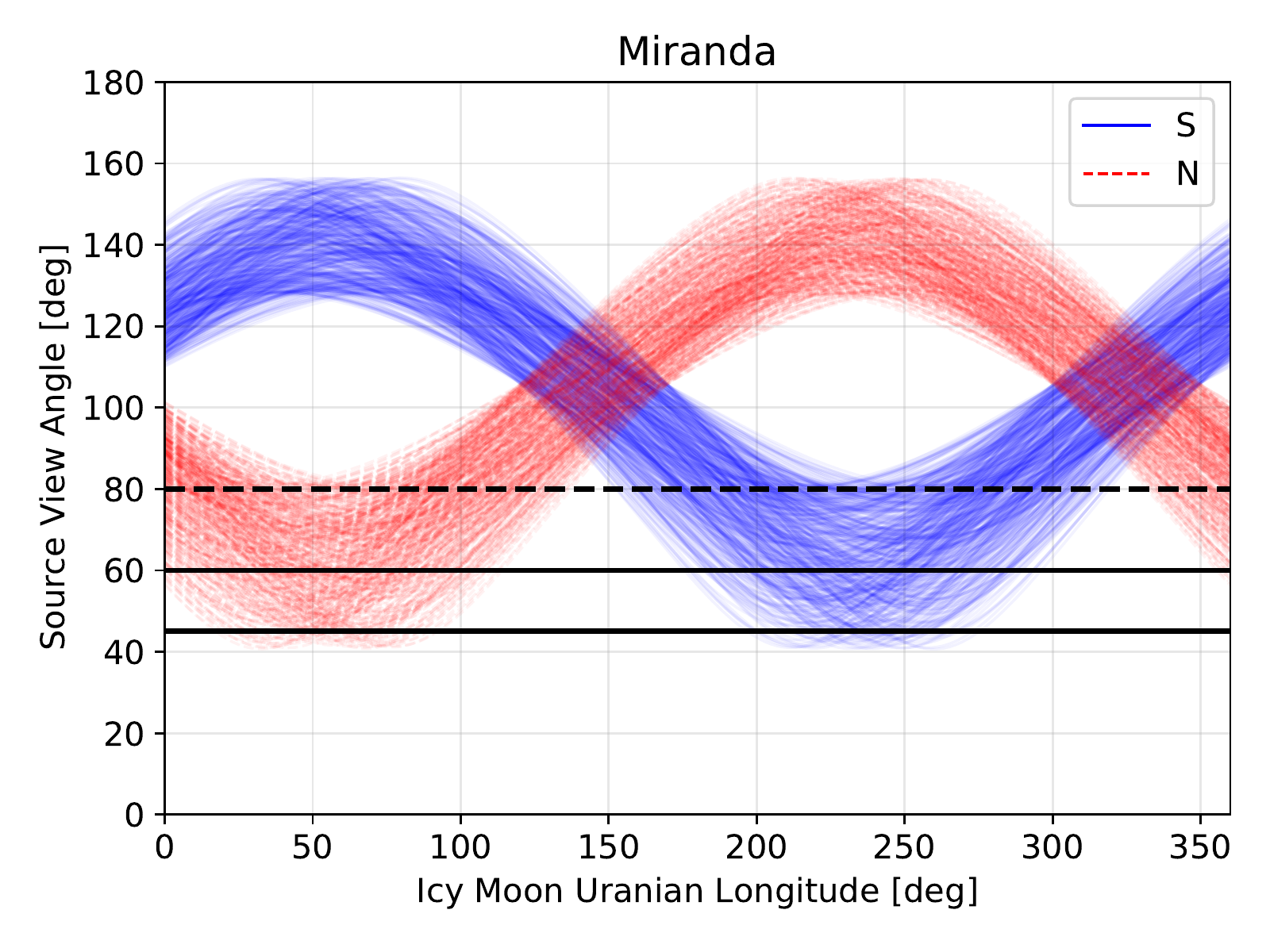}
\noindent\includegraphics[width=0.49\textwidth]{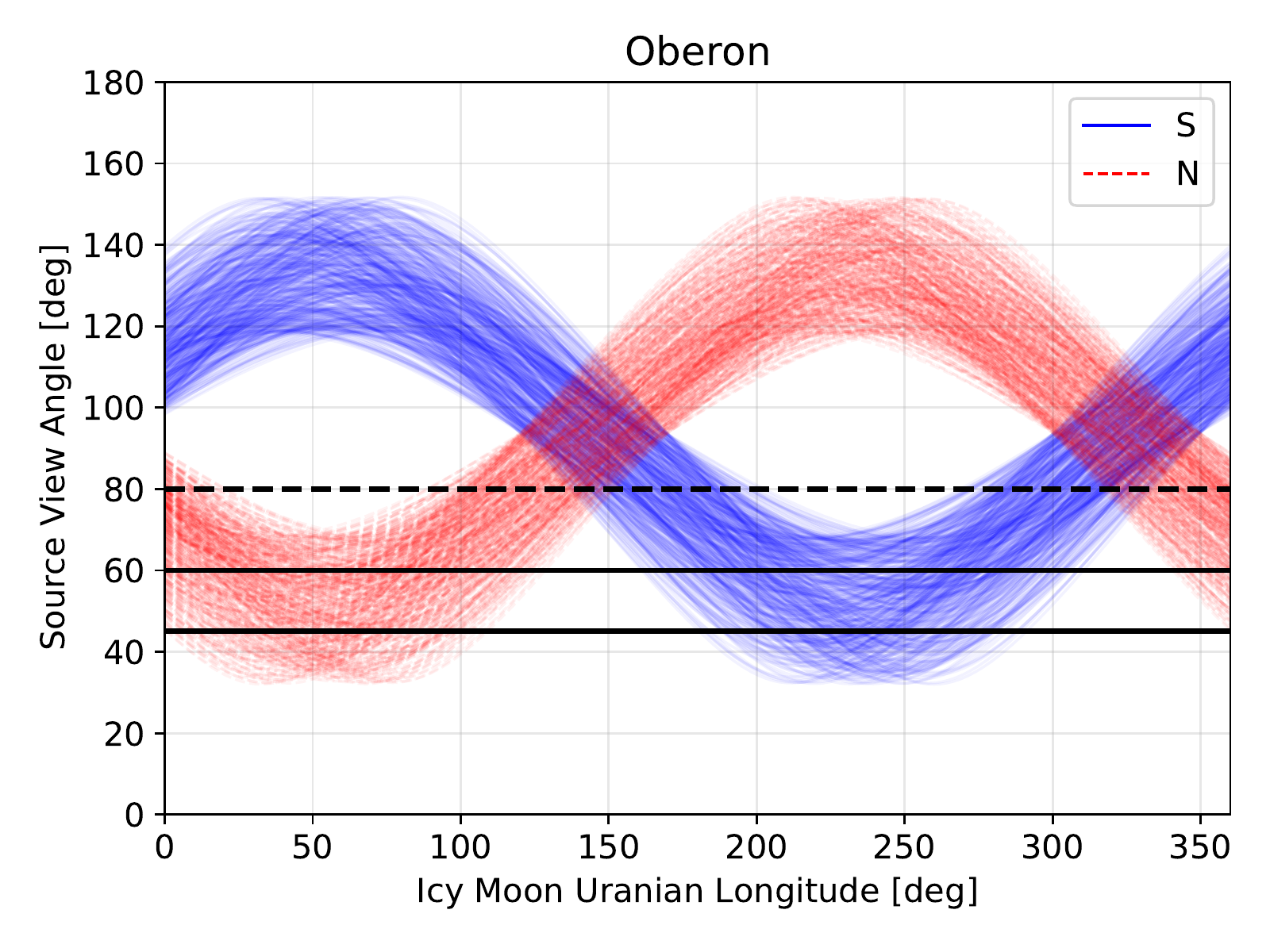}
\caption{Source view angle ($\theta_\mathrm{view}$ as defined in Figure~\ref{fig:geometry}) as a function of the icy moon's Uranian Longitude. The traces correspond to points $\mathbf{r}_\mathrm{src}$ sampled over the source extent of the southern source (blue traces) and for a northern source (red traces). The southern UKR source is modelled based on~\cite{Menietti_1990}. The northern source was not observable by Voyager-2 and, as a proxy, we have included it as a copy of the southern source model mapped to the antipodal region. Modelling the radio emission of the northern source is future work (see \S\ref{sec:discussion} for further discussion). The solid horizontal lines corresponds to the view angles where the southern UKR source would illuminate the icy moon. The dashed black line corresponds to the theoretical maximum cone opening angle from~\citeA{Menietti_1990}. The closest (Miranda) and farthest (Oberon) of the icy moons of interest are shown to illustrate the extremes. }
\label{fig:illumination_view_angles}
\end{figure}

In Figure~\ref{fig:illumination_view_angles} we show the southern UKR source view angle $\theta_\mathrm{view}$ with respect to the icy moons Miranda and Oberon. The source will illuminate the icy moon when it is $\pm80^\circ$ away from the Uranian longitude of the centroid of the southern UKR source at $\sim235^\circ$. 
The northern radio source was not observable by Voyager-2 so we do not have information on its size and beaming properties. As a proxy, we have also included the northern source as a clone of the southern source located rotated to the antipodal point. While it is expected that there are differences between the northern and southern sources, this is meant to show the level of source availability expected. See \S\ref{sec:discussion} for a more detailed discussion. 

\section{Receiver Model}
\label{sec:receiver}
\subsection{General Properties}
We use the Cassini Radio Plasma Wave Science (RPWS) instrument~\cite{Gurnett_2004} with a modified digitizer (1 MHz instantaneous bandwidth) and signal processing chain as a baseline for this study. The key properties are the antenna sensitivity and noise contributions in the environment of the Uranian icy moons. 

The sensitivity of the instrument is determined by a combination of antenna effective length and instrument noise. In the frequencies of interest ($< 1$~MHz), the electrically short antenna approximation is valid. The dipole has an effective length $L_\mathrm{eff}\simeq 3.1$~m including stray capacitance losses~\cite{Zarka_2004b} but a physical length of 7.3~m (tip-to-tip). The noise contributions (internal and external) are covered in the next subsection. 

\subsection{Radio Frequency Noise}
\label{sec:noise}
\begin{figure}[ht]
\centering
\noindent\includegraphics[width=0.8\textwidth]{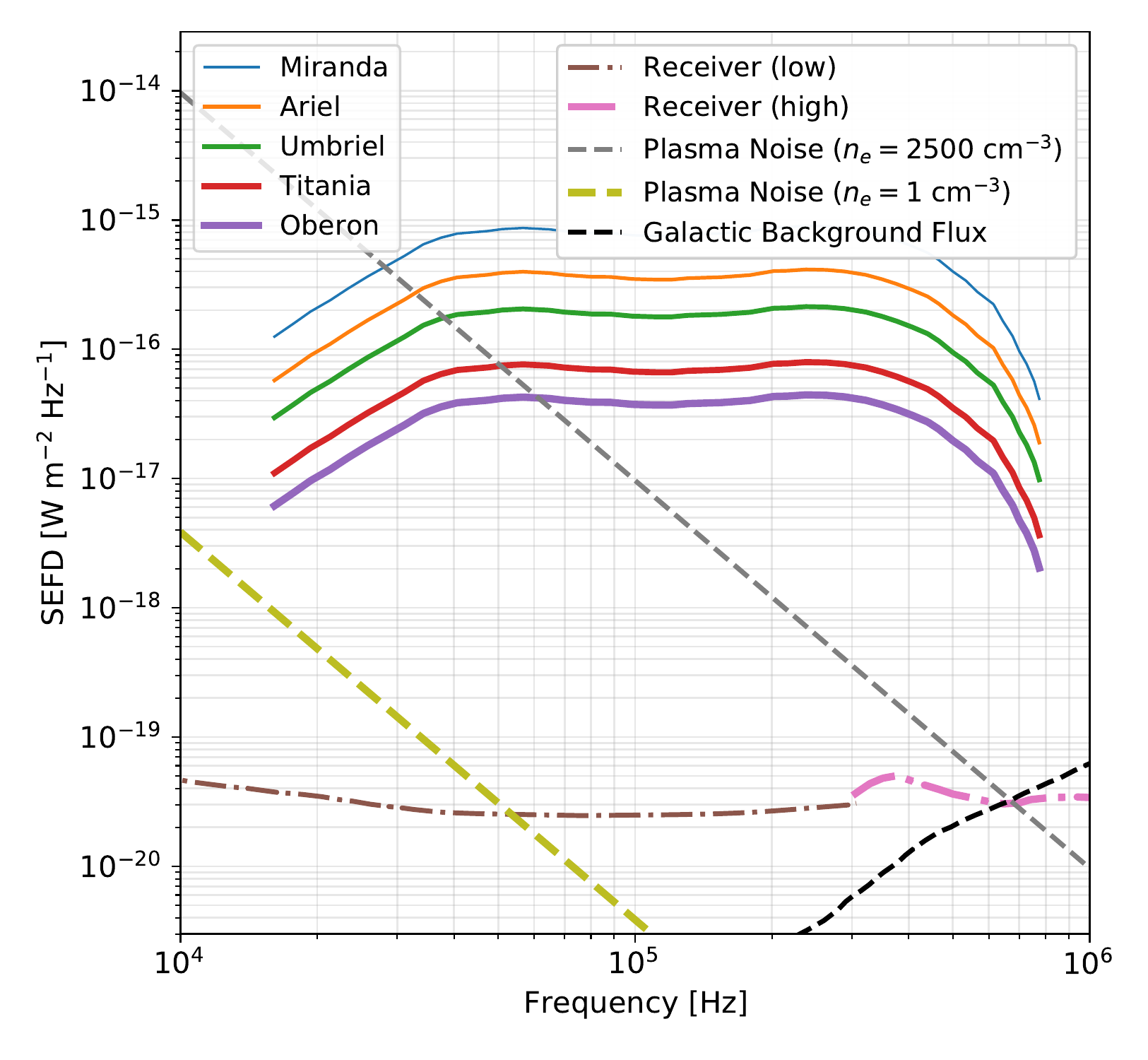}
\caption{Noise backgrounds in spectral equivalent flux density (SEFD) compared to expected flux densities of UKR at the icy moons. The fluxes at each icy moon of interest is shown using solid colored lines. The receiver noise is the Cassini low and high frequency band receiver measured prior to antenna deployment as reported in~\citeA{Zarka_2004b} is shown in dash-dotted lines. The lower bound on plasma noise corresponding to an electron density of $n_e=1.0 \ \mathrm{cm^{-3}}$ and temperature $T_e=3\times 10^3$~Kelvin is shown with a yellow dashed line and the upper bound corresponding to $n_e=2500 \ \mathrm{cm^{-3}}$ and temperature $T_e=10^3$~Kelvin is shown with a gray dashed line (see text for details on the choice of parameters). The Galactic background flux is shown as a dashed black line. }
\label{fig:noise}
\end{figure}

A noise calibration of the Cassini RPWS is provided by~\citeA{Zarka_2004b}. The internal receiver noise is estimated by taking power spectral density measurements with the antennas stowed prior to deployment. Using the effective length of the dipole antennas, we have converted these data to spectral equivalent flux density (SEFD) as shown in Figure~\ref{fig:noise}. The conversion between power at the receiver (in units of $\mathrm{V^2  \ Hz^{-1}}$) to flux  (in units of $\mathrm{W  \ m^{-2}  \ Hz^{-1}}$) is given by $K=Z_0 L^2_\mathrm{eff}\simeq3530 \mathrm{\ m^2 \Omega}$ where $Z_0$ is the impedance of free space and $L_\mathrm{eff}$ is the effective length of the antenna referenced at the receiver including stray capacitance losses (\citeA{Zarka_2004b}). The figure also includes the flux of the UKR emissions at Miranda and Oberon and are more than three orders of magnitude greater than the receiver noise. The Galactic background noise from~\citeA{Manning_2001} is also shown in Figure~\ref{fig:noise} and is below the receiver noise except between 600~kHz -- 1~MHz where it becomes comparable to the receiver noise. 

The plasma noise dominating at lower frequencies (Figure~\ref{fig:noise}) is due to the currents induced on the antenna by the random motion of free electrons in its immediate vicinity. The plasma noise induced at the terminals of the antenna depends on the half-length of the dipole $L_{1/2}$, the number density of electrons $n_e$, and their temperature $T_e$. The equation below is adapted from~\citeA{Meyer-Vernet_1989} with scale factors relevant to this concept:

\begin{equation}
    \langle V^2_\mathrm{plasma} \rangle \simeq 
    4.1\times10^{-17} \ 
    \mathrm{\frac{V^2}{Hz}} \
    \left( \frac{n_e}{\mathrm{1 \ cm^{-3}}} \right)
    \left( \frac{T_e}{\mathrm{3\times10^3 \ K}} \right)
    \left( \frac{f}{\mathrm{100 \ kHz}} \right)^{-3}
    \left( \frac{L_{1/2}}{\mathrm{3.65 \ m}} \right)^{-1}
\end{equation}

In terms of system-equivalent flux density (SEFD), the plasma noise is given by 

\begin{equation}
    \langle \mathrm{SEFD}_\mathrm{plasma} \rangle \simeq 
    1.2\times10^{-20} \ 
    \mathrm{\frac{W}{m^2 \ Hz}} \
    \left( \frac{n_e}{\mathrm{1 \ cm^{-3}}} \right)
    \left( \frac{T_e}{\mathrm{3\times10^3 \ K}} \right)
    \left( \frac{f}{\mathrm{100 \ kHz}} \right)^{-3}
    \left( \frac{L_{1/2}}{\mathrm{3.65 \ m}} \right)^{-1}
\label{eq:sefd_plasma}
\end{equation}

Since no data is available on the electron density and temperature near the surface of Uranian icy moons we provide a lower and upper bound. For the lower bound, we use measurements of plasma in the vicinity of Uranus, but far from any moons, taken with Voyager-2~\cite{Sittler_1987}. The closest approach of Voyager-2 to Miranda, Ariel, Umbriel, Titania, and Oberon was 29,000~km, 127,000~km, 325,000~km 365,200~km 470,600~km, respectively~\cite{Stone_1987} while ionospheric scale heights are expected to be $<1,000$~km. The plasma electron temperature during this pass was typically $T_e\simeq3\times10^3$~eV while the plasma electron density was typically $n_e\simeq 10^{-3}$~cm$^{-3}$ but could go as high as $n_e\simeq 1$~cm$^{-3}$. The expected lower bound plasma noise level shown in Figure~\ref{fig:noise} uses $T_e=3\times10^3$~eV and $n_e = 1.0$~cm$^{-3}$. This plasma electron density and temperature values do not result in a significant source of noise for most of the band of interest. 

\begin{figure}
    \centering\includegraphics[width=0.8\textwidth]{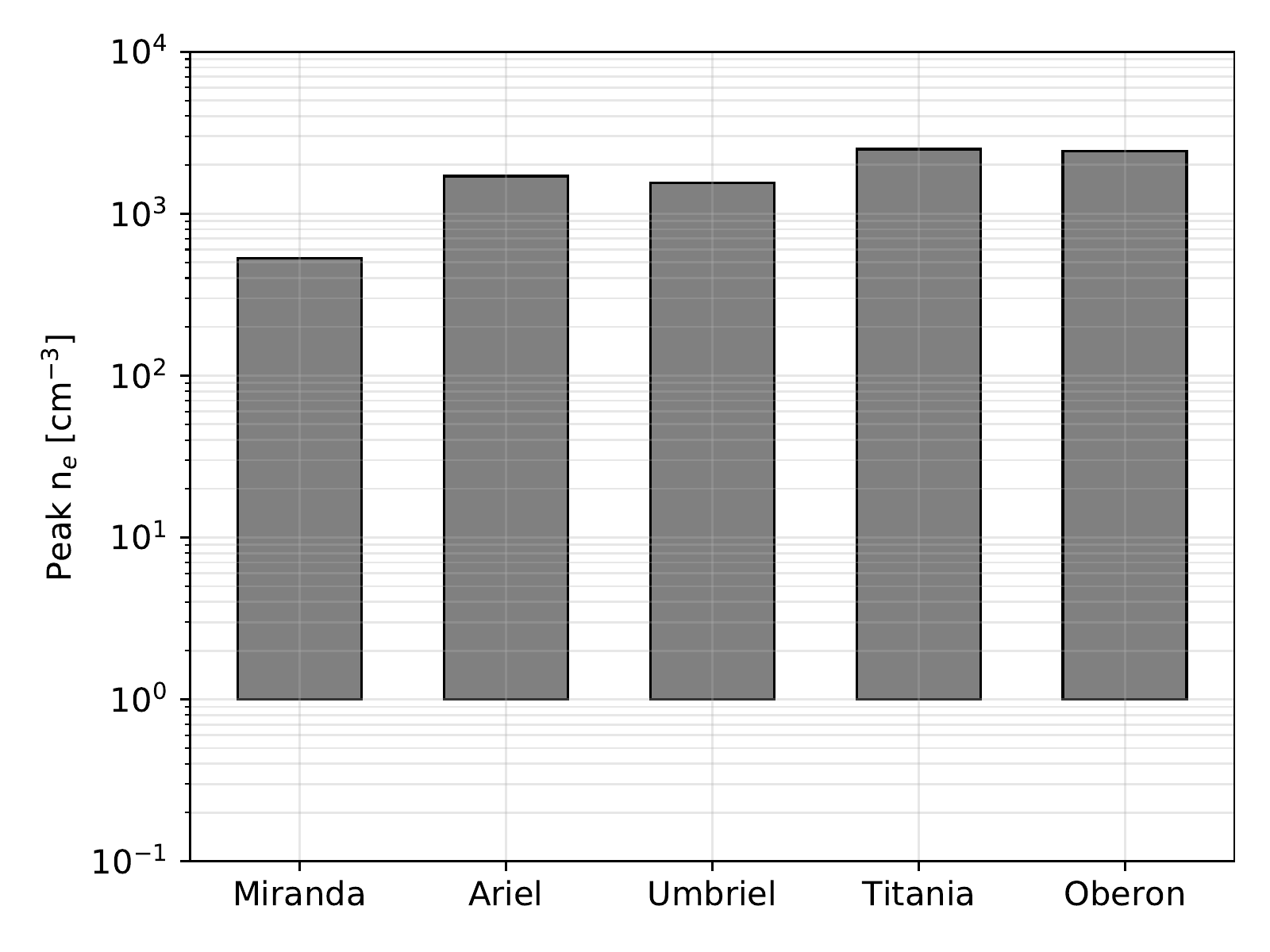}
    \caption{Range of possible values for the peak ionospheric electron density of the Uranian icy moons. The lower bounds are from Voyager-2 measurements of the plasma density in the Uranian system. The upper bound is obtained by scaling to the peak electron density and surface gravity of Europa. These upper limits are aggressively pessimistic given that, unlike Europa, Uranian icy moons do not reside in a plasma torus nor are they expected to be active.} 
    \label{fig:plasma_density_UL}
    \end{figure}

For the upper bound, we can estimate the electron plasma density $n_e$ assuming its ratio to surface gravity $g_\mathrm{surf}$ is approximately constant. The peak electron density of Europa's ionosphere during daytime conditions is $n_{e, Eu}\simeq10^{4}~\mathrm{cm^{-3}}$ and drops to levels consistent with zero during nighttime~\cite{Kliore_1997}. The surface gravity of Uranian icy moons ranges from $7.9\times10^{-3}g$ (Miranda) to $3.7\times10^{-2}g$ (Oberon), where $g$ is the surface gravity of Earth, compared to Europa with $1.3\times10^{-1}g$. The scaled peak ionospheric density of Uranian icy moons is given by assuming the ratio of peak electron density and surface gravity $n_\mathrm{e, peak}/g_\mathrm{surf}$ is constant. The upper bound in peak plasma density derived in this manner are shown in Figure~\ref{fig:plasma_density_UL}. The resulting plasma noise profile, assuming an electron temperature $T_e\sim10^3$~K which bounds the atmospheric temperature of Europa typically assumed to be in the hundreds of Kelvin~\cite{Kliore_1997}, is shown in Figure~\ref{fig:noise} with the curve labeled Plasma Noise ($n_\mathrm{e} \ = \ 2500 \ \mathrm{cm^{-3}}$). Note that this bound is aggressively pessimistic since the icy moons of Uranus, unlike Europa, do not reside in a plasma torus and are not expected to be active. 

\section{Target Properties}
\label{sec:target}
We consider the interior structure and composition models by \citeA{Castillo2023} to evaluate the potential to reveal the interior structure of the Uranian satellites using passive radar. The ice shells of the Uranian satellites differ significantly from the ice shells that have been previously studied for radar sounding, e.g., Europa \cite{Kalousova2017} and Enceladus \cite{Soucek2023}. The ice shells of all major Uranian satellites are generally assumed to be too cold for convective heat transfer to be operating at present \cite{bierson2022, Hussmann2006} and with thicknesses on the order of 120 to 300 km \cite{Castillo2023}. Based on carbonaceous chondrite composition supported by ground based infrared spectroscopy, the satellites could be rich in nitrogen-bearing species \cite{cartwright2020,cartwright2023}. Furthermore, the presence of subsurface oceans could imply high porosity in the upper crust providing increased thermal insulation. Porosity might have two origins: primordial microporosity (accreted material) and macroporosity introduced by large impacts. 

While cold ice tends to be very transparent to radio waves, attenuation increases with temperature and is also affected by impurities, specifically those that are soluble in the ice lattice (e.g., Cl$^-$, NH$_4^+$, H$^+$). Importantly, this implies that attenuation increases as an ice-ocean interface is approached. In addition, the porous crust could lead to volume scattering. However, due to the long wavelength, surface roughness losses are expected to be negligible. Therefore, only attenuation and volume scattering are investigated in the following. For this purpose we consider the following end-member models derived from \citeA{Castillo2023} for Ariel/Umbriel and Titania/Oberon. Both pairs of moons are expected to be similar enough in structure and composition to be treated together. A sharp ice-ocean interface is likely to be highly reflective. Using sea ice brines as an analog~\cite{stogryn1985}, we predict a reflection coefficient of $>-0.1$ dB for an ice-ocean interface frequencies between 10 kHz and 1 MHz. Miranda is not considered here as it is not expected to have an ocean but we will discuss the potential detection of an ice-rock interface in Section \ref{sec:outcomes}.


\subsection{Ice Shell Model and Radio Frequency Attenuation}
\label{sec:attenuation}
To model the attenuation in ice we assume a conductive temperature profile with a surface temperature of $T_s$ = 70~K and two ocean cases. One with a thin ocean, highly enriched in ammonia and with an equilibrium temperature at the ice ocean interface at depth $b$ of  $T_b$ = 180~K and a second case with a thick ocean and a temperature of $T_b$ = 268~K. The structural and compositional parameters are summarized in Table \ref{tab:structure}. The temperature as a function of depth $z$ is represented by the equilibrium profile for a thermally conductive ice shell:
\begin{equation}
T(z) = T_s \exp\left(z\frac{\ln(T_b/T_s)}{b}\right)
\end{equation}

The attenuation in ice depends on the electrical conductivity of the material which, in addition to the temperature, further depends on the concentration of lattice soluble impurities. Using the model and the parameters of \citeA{MacGregor2015}, the conductivity of pure ice as a function of frequency can be approximated by
\begin{equation}
    \sigma_p = \omega\epsilon_0\mathfrak{Im}\left(\frac{\Delta\epsilon'}{1+(i\omega\tau)^{1-\alpha}}\right)\,,
\end{equation}
with the angular frequency $\omega$, the permittivity in vacuum $\epsilon_0$, the dielectric susceptibility $\Delta\epsilon$, the relaxation time $\tau$, and the Cole-Cole distribution parameter $\alpha$=0.1 \cite{MacGregor2015}. In the presence of impurities, the conductivity becomes
\begin{equation}
\sigma = \sigma_p\exp\left[\frac{E_{ice}}{k_b}\left(\frac{1}{T_r}-\frac{1}{T}\right)\right]+\sum_i^N\mu_{i}M_i\exp\left[\frac{E_i}{k_b}\left(\frac{1}{T_{r,i}}-\frac{1}{T}\right)\right] \,.
\end{equation}
 The in ice 2-way attenuation as a function of depth is then given by
\begin{equation}
A_{2}=2\frac{10\log_{10}(e)}{10000\epsilon_0\sqrt{\epsilon_{ice}}c}\int_0^b\sigma(z)dz \,.
\end{equation}
We derived the ice shell composition from the ocean composition assuming that the impurities in the ice follow a partition coefficient of 0.137 for Cl in presence of ammonium \cite{gross1977concentration} for equilibrium freezing.

\begin{table}[t!]
    \centering
    \begin{tabular}{c|cc|cc}
    & \multicolumn{2}{c|}{Ariel/Umbriel} & \multicolumn{2}{c}{Titania/Oberon}\\
    \hline
    Moon Radius [km] & \multicolumn{2}{c|}{580} & \multicolumn{2}{c}{770}\\
    H$_2$O Layer  [km] & \multicolumn{2}{c|}{190} & \multicolumn{2}{c}{240}\\
    \hline
    & Thin Ocean & Thick Ocean & Thin Ocean & Thick Ocean \\
    \hline
    Ocean Thickness [km] & 2 & 25 & 4 & 50 \\ 
    Ocean Cl [Mol/kg] & 4 & 0.5 & 3 & 0.1 \\ 
    Ocean NH$_3$ [Mol/kg] & 20 & 5 & 9 & 1 \\   
    Ocean NH$_4$ [Mol/kg] & 4 & 0.75 & 3 & 0.9 \\   
    \end{tabular}
    \caption{Structural and composition models for the attenuation model. For each moon pair we consider a thin ocean case and a thick ocean case. Parameters derived from \citeA{Castillo2023}.}
    \label{tab:structure}
\end{table}

\begin{figure}[t!]
\noindent\includegraphics[width=\textwidth]{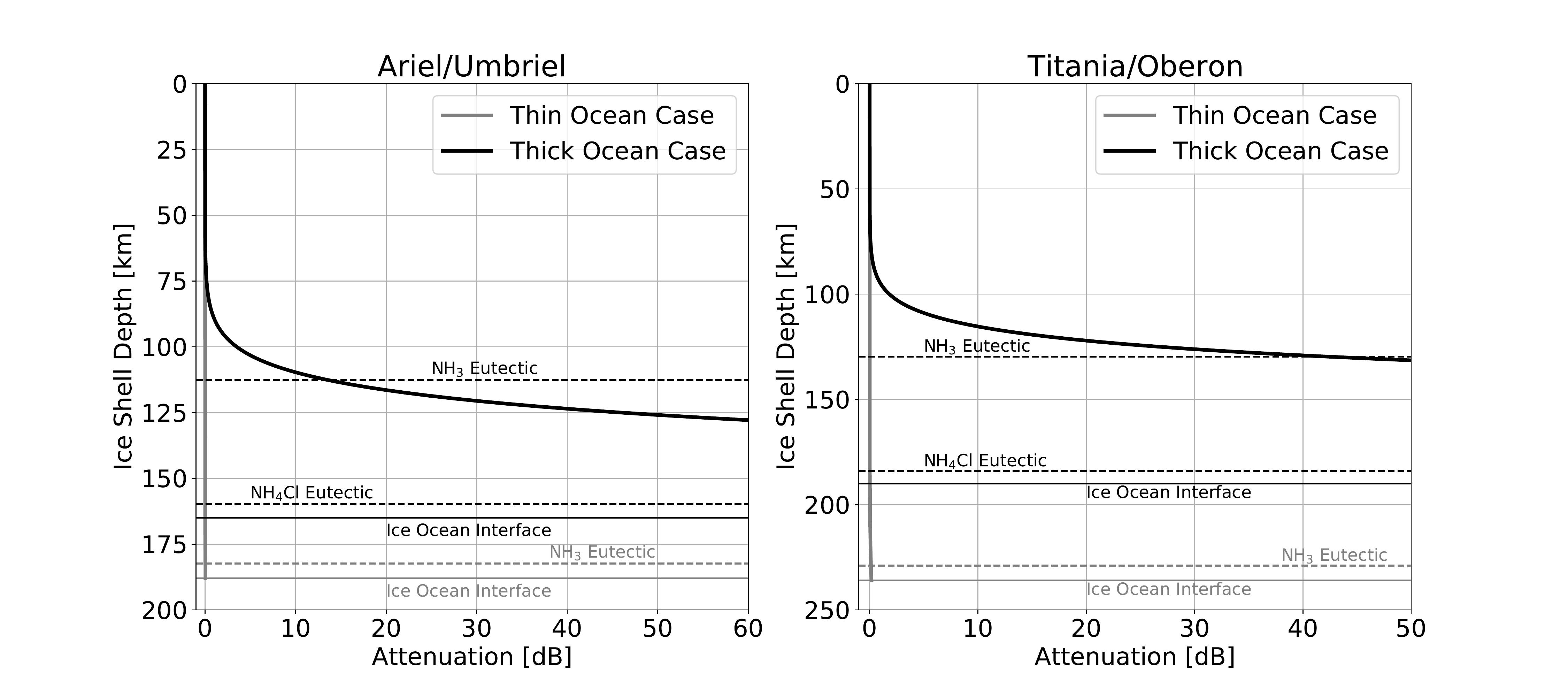}
\caption{Radar attenuation as a function of ice shell depth for Ariel and Titania. The results for Ariel are assumed to be identical to Umbriel, and Titania to Oberon, respectively, due to the similar interior structure of the two moons. Shown are the results for a thick ocean case and a thin ocean case, with the respective locations of the ice-ocean interfaces and the eutectic temperatures for compositionally relevant aqueous solutions.}
\label{fig:attenuation} 
\end{figure}

The resulting 2-way attenuation as a function of depth is shown in Figure \ref{fig:attenuation} and calculated for a center frequency of 100 kHz, however the frequency dependence of ice conductivity is relatively flat therefore the changes on the results for different frequencies between 100 kHz and 1 MHz are rather subtle for temperatures above -55 $^\circ$C but tend to decrease with lower temperatures \cite{fujino1967}. Due to the similarities in interior structure, we grouped the parameters and results for Ariel and Umbriel, and for Titania and Oberon considering a thin ocean and thick ocean case for each moon pair as described by the parameters given in Table \ref{tab:structure}. The best case scenario in terms of direct ocean detection would be an ocean at the eutectic point which would lead to a very cold ice-ocean interface and therefore an equally cold ice-shell. In such a scenario the attenuation would be effectively negligible. 

In case of a thick ocean, the warm ice close to the ice-ocean interface in combination with the elevated concentration of impurities within the shell would lead to high attenuation within the ice. This situation would make it unlikely to directly detect the ocean. However, attenuation only becomes significant below the depth where the temperature is above the NH$_3$ eutectic temperature, referred to here as the eutectic interface. Below the eutectic interface, the ice is partially molten, where the amount of melt stable is governed by the temperature and concentration of impurities in the ice \cite{Wolfenbarger2022}. The detection of a eutectic interface would provide a constraint on the temperature profile of the ice shell and, if the composition is known, on the  location of the putative subsurface ocean. Similar hypotheses have been formulated for the use of active radar sounding in the context of Europa \cite{Culha2020} and Enceladus \cite{Soucek2023}.

\subsection{Volume Scattering}
\label{sec:scattering}
While large porosities are unlikely for larger moons as the porosity significantly decreases above pressures of 25 MPa within the Uranian satellites \cite{Castillo2023}, there could still be a porous outer crust resulting from primordial microporosity and fracturing events. Increased porosity values can lead to significant scattering losses if the pore sizes are large compared to the radar wavelength (see discussion within \citeA{Eluszkiewicz2004} and \citeA{Aglyamov2017} for Europa). In the case of the kilometric radiation from Uranus however, the long wavelength significantly reduces the susceptibility to volume scattering. The effect from Mie-scattering can be estimated the using the anomalous diffraction approximation of 
\citeA{hulst1981}. Note that this approximation is assuming large spheres compared to the wavelength and tends to overestimate scattering losses for lower frequencies, so can be assumed to be conservative in our case. The scattering efficiency factor in this approximation is given by
\begin{equation}
    Q = 2 - \frac{4}{p}\sin(p)+\frac{4}{p^2}(1-\cos(p))\,,
\end{equation}
with 
\begin{equation}
    p = 4\pi r \frac{(n-1)}{\lambda}\,.
\end{equation}
In the equation above, $r$ is the radius of the spheres, $\lambda$ the radio wavelength, and $n$ the ratio of refractive indices. With the efficiency factor $Q$ we can calculate the optical depth of the ice with total thickness $d$ and porosity $\phi$ in the same way as \citeA{Aglyamov2017} as
\begin{equation}
    \tau = \frac{3\phi d}{4r}Q\,,
\end{equation}
and the two-way scattering losses by $L = \exp\left(-2\tau\right)$. Using the extremely conservative case of a porosity of 30\% ($\phi=0.3$) for the entire ice shell of $d=180$ km thickness with sphere radii of $r=5$ m, $n=1.75$, $\lambda=3$ km, we find scattering losses of less than 9 dB for the entire ice shell. Therefore, we conclude that volume scattering is not an obstacle for the proposed technique.

\subsection{Passive Signal-to-Noise Ratio}
\label{sec:SNR}
\begin{figure}[ht]
\centering
\noindent\includegraphics[width=0.6\textwidth]{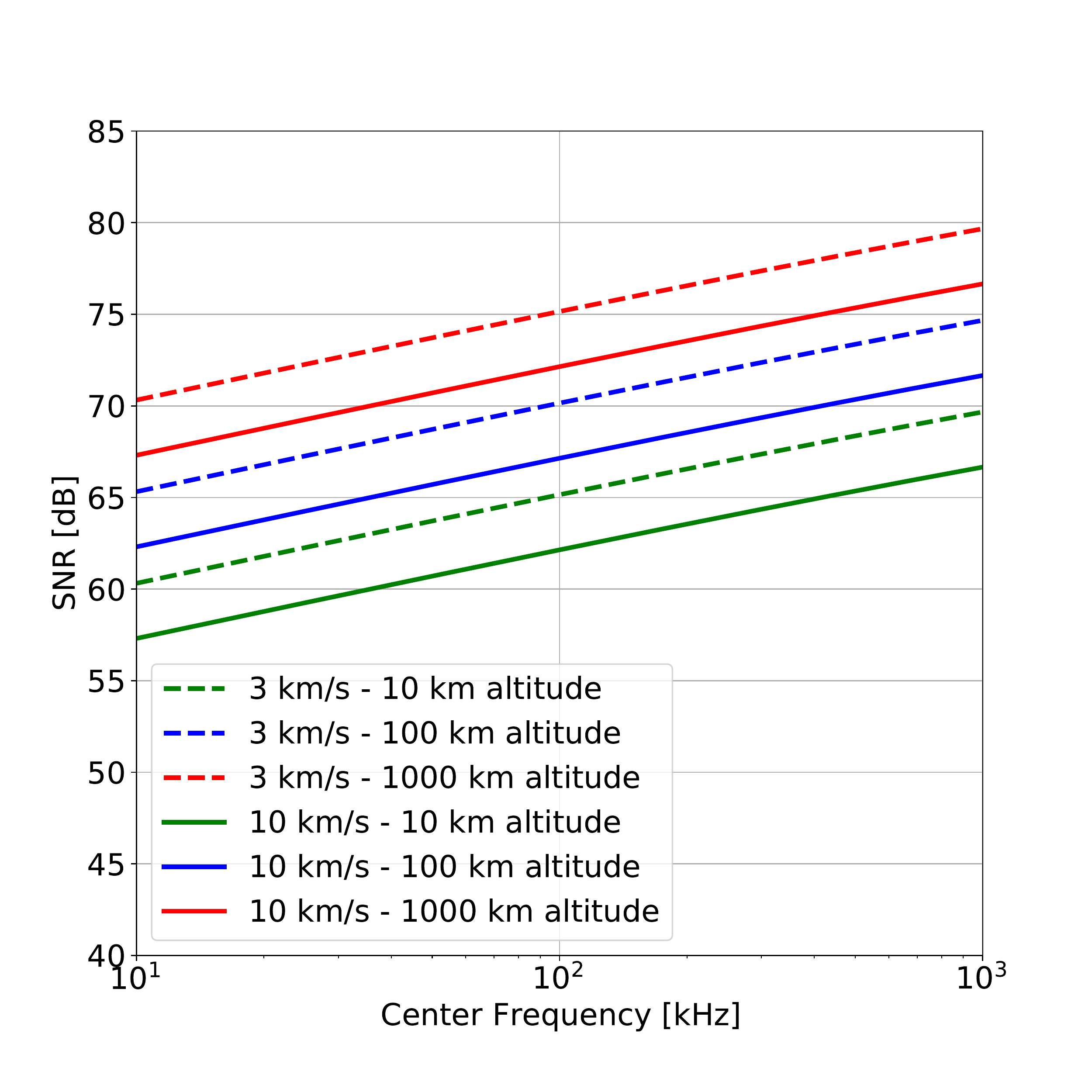}
\caption{Passive signal to noise ratio for the surface reflection as a function of center frequency, altitude and flyby speed. In all scenarios we expect to obtain around 55 - 80 dB.}
\label{fig:snr}
\end{figure}

For the calculation of the passive sounding Signal-to-Noise Ratio (SNR) we follow the approach of \citeA{Schroeder_2016}. In the context of passive sounding, this term can be ambiguous as, by definition, the noise is the signal. Therefore, this value should be understood as the strength of the auto-correlated signal versus the UKR background. Other sources such as the Galactic background are not included in the following calculation. Further, we only calculate the surface SNR for a perfectly reflecting interface. This number should be compared against the estimated attenuation and bulk scattering losses described in Section \ref{sec:attenuation} and \ref{sec:scattering}. When the source being used for passive sounding is significantly larger than other backgrounds, the passive SNR then generally depends on how much of the noise from the source can be integrated. Therefore, not only the altitude $h$ but also the flyby speed $v$ affect the SNR. Further, the higher bandwidths $\beta$ are favorable. Here, we assume that the bandwidth is half the center frequency, which will lead to higher SNR's for higher frequencies \cite{Schroeder_2016}.
\begin{equation}
\mathrm{SNR} = \frac{2\sqrt{h\lambda}\beta}{v\left(1+\sqrt{\frac{h}{\lambda}}\tan(\sigma_s)\right)^2}
\end{equation}
Here, $\sigma_s$ is the surface slope at the wavelength scale. Assuming a fractal surface, the slope at these scales is expected to be small therefore the associated term is negligible. For the flyby groundspeed we consider two end-members with 3 km/s on the lower end and 10 km/s on the upper end. Based on the maximum altitudes inferred in Section \ref{sec:extent}, we consider 10, 100, and 1000 km. The results are shown in Figure \ref{fig:snr} and suggest 55 - 70 dB for 100 kHz and 65 - 80 dB for 1 MHz.

\section{Expected Return Signal Characteristics}
\label{sec:outcomes}
Based on the discussion of the target properties in Section \ref{sec:target}, we can hypothesize a set of interior structure scenarios and their predicted signature from passive radar sounding. 

A passive radar operating at kilometric wavelength has to compromise in terms of vertical resolution. Further, integration times over the groundtrack have to be balanced against the horizontal resolution, especially when compared to actively pulsed radars operating at MHz frequencies. We estimate that over the course of a flyby, only a few range lines can be recorded. Ultimately the number will be a trade-off between the horizontal resolution and the SNR.   

Given the dominant effect of ocean temperature on the attenuation and the similarities of the results for the individual moons, we can expect three plausible cases: The presence of no ocean, the presence of a cold ocean, or the presence of a warm ocean. In all three scenarios, some interface will likely be detected but the characteristics of the return signal would be different. The cold ($<200$ K) ocean case should return a signal from the ice-ocean interface exceeding the strength of the surface return. This is due to the low scattering and attenuation losses on one hand, and the strong reflection coefficient of liquid water on the other hand. This scenario would enable a direct, unambiguous ocean detection and simultaneously determine the thickness of the overlaying ice shell. Further, the ratio of the amplitudes of the surface return and ocean return are informing the attenuation and therefore constrain the temperature and composition of the ice shell. 

In case of a warm ($>200$ K) ocean, the attenuation is likely too strong to allow the direct detection of an ice-ocean interface. As the ocean extent is assumed to be small, the high concentration of impurities in the lower ice layers will make ice probing by radio waves challenging. However, also in that scenario the NH$_3$ eutectic interface could be probed with less than 50 dB of attenuation on Titania and Oberon and less than 20 dB of attenuation on Ariel and Umbriel. The eutectic would constitute the first liquid interface and presence of liquids would likely shadow the structure beneath. In this scenario, the use of passive radar would therefore be most powerful in combination with a magnetometer, which could detect an induction signal in the warm ocean case \cite{Cochrane2021}.

In the case that no ocean is present, passive radar would likely still detect the ice-mantle interface as the ice shell is expected to be cold in this case \cite{Castillo2023}. As the return signal is expected to be less strong than in the cold ocean case, there is some ambiguity from one return alone as a dim return could also originate from a somewhat warm ocean (due to the enhanced attenuation as the ice-ocean interface is approached). Having multiple range lines distributed over the ground track could characterize the interface and help discriminating an ocean return from a mantle return. Further ways to discriminate between the two cases would be to test if an induced magnetic field is absent or obtain constraints on the shell temperature, for example by performing UKR occultation measurements to probe the attenuation profile of the ice shell (see \S\ref{sec:discussion}).  

\section{Discussion}
\label{sec:discussion}

    This study is focused on a first evaluation of the feasibility of passive sounding for subsurface oceans in the icy moons of Uranus using UKR emissions. Passive radar sounding presents a complementary technique to magnetic induction; the low electrical conductivity of a cold, ammonia-rich ocean that challenges magnetic induction measurements is favorable for sounding the ice-ocean interface while extended source size and radio beam patterns limit access to the closer moons. Although this technique is promising, there are a number of modeling aspects that need to be refined in order to minimize the risks of a future implementation. In this section, we discuss some of the developments needed. Their quantification fall outside the scope of this paper and will be the subject of future work.


    {\it Northern UKR source:} The Voyager-2 flyby of Uranus only partially observed the southern source and none of the northern source. While the northern and southern kilometric radio sources in well-studied gas giants (Jupiter and Saturn) are similar, they do show differences in frequency cutoff and potentially also in size. The uncertainties in source size and radio emission beam impact source availability and maximum altitude for passive sounding, which are key parameters for planning flybys. These uncertainties can be further characterized and potentially reduced by using forward-modelling computational tools such as the Exoplanetary and Planetary Radio Emission Simulator (ExPRES)~\cite{Louis_2019}. This simulation can take a magnetic field models of Uranus, of which there are many possibilities (see \citeA{podolak1991}), along with a plasma density model to predict the visibility of radio emissions. These models can be tested against Voyager-2 data for the southern UKR source and applied to characterize the uncertainties in the northern UKR source. ExPRES also allows an auroral oval model as input to predict visibility of radio emissions. Ultraviolet observations of the Uranian aurorae with the Hubble space telescope \cite{balcerak2012} could be applied as additional input for these predictions. 


    {\it Solar radio bursts:} Solar radio bursts could interfere with a passive sounding flyby. We can bound the probability that this occurs via Equation~\ref{eqn:p_solar_burst}. 
    
    \begin{equation}
    P_{SB} < 0.015
    \left(\frac{R_\mathrm{III}}{6.6 \ \mathrm{day^{-1}}}\right)
    \left(\frac{T_{100\mathrm{kHz}}}{1 \ \mathrm{hr}}\right)
    \left(\frac{P(>10^{-18} \ \mathrm{W \ m^{-2} \ Hz^{-1}})}{0.055}\right)
    \label{eqn:p_solar_burst}
    \end{equation}
    
    The rate of Type III bursts is $R_\mathrm{III}\sim6.6$ per day at solar maximum and decreases by approximately an order of magnitude at solar minimum~\cite{Ndacyayisenga_2021}. We do not consider type II bursts since they are more than an order of magnitude less frequent than type III bursts at frequencies $<1$~MHz, and generally much weaker in signal strength~\cite{Krupar_2018}. An icy moon flyby lasts for the order of minutes to tens of minutes (see \S\ref{sec:SNR}) compared to the $\sim1$~hour duration of type III bursts at 100~kHz so we scale by the duration of the radio burst $T_{100\mathrm{kHz}}\simeq1 \ \mathrm{hr}$. Finally, we weigh in the probability that the burst exceeds a flux density of $10^{-18} \ \mathrm{W \ m^{-2} \ Hz^{-1}}$ at the Uranian system, which is conservatively chosen to be roughly an order of magnitude below the UKR flux at Oberon. The probability of this $P(>10^{-18} \ \mathrm{W \ m^{-2} \ Hz^{-1}})\simeq 0.055$ is based on~\citeA{Krupar_2018}, where we have scaled by the square of the distance between Earth and Uranus. These conservative estimates result in a probability of a type III Solar Radio Bursts $P_{SB}$ smaller than 1.5\% making it a negligible concern.

    {\it Icy moon ionospheres:} The ionosphere of Uranian icy moons is not well constrained and can limit the minimum usable frequency for sounding, affect the signal shape via frequency-dependent dispersion, and result in additional losses due to Faraday rotation induced by interaction with the Uranian magnetic field. 
    
\begin{figure}
    \centering\includegraphics[width=0.8\textwidth]{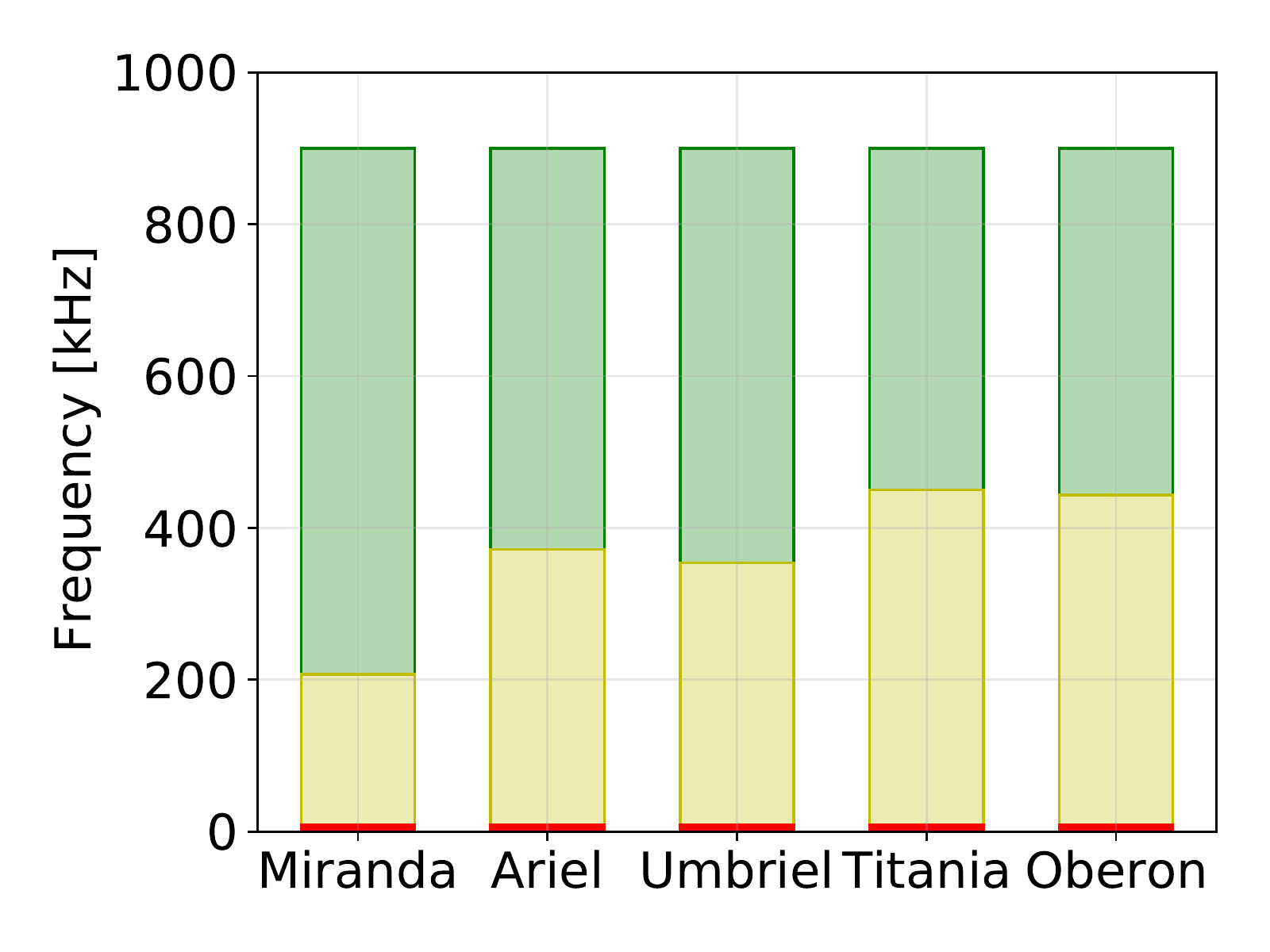}
    \caption{Frequency band available for passive sounding for each icy moon. The upper value of 900~kHz is due to the cutoff frequency of the UKR source. The green bar extends down to the plasma frequency upper limit derived from scaling to Europa's peak ionospheric electron densities ( Figure~\ref{fig:plasma_density_UL}) and scale heights. These upper limits are aggressively pessimistic given that, unlike Europa, Uranian icy moons do not reside in a plasma torus nor are they expected to be active. The yellow bars cover the uncertain range between the plasma frequency upper limit and the ambient plasma frequency in the Uranian system ($f\simeq 9~kHz$). Even with these pessimistic upper bounds, a significant portion of the spectrum of UKR emissions penetrate through the icy moon ionospheres and allows for passive sounding.}
    \label{fig:plasma_freq_UL}
    \end{figure}

    The peak electron density of the ionosphere determines the cutoff frequency below which radio emissions will not propagate into the surface or subsurface of the icy moon. The  ionospheric cutoff frequency, below which radio signals will not propagate, is determined by the plasma frequency 
    \begin{equation}
    f_\mathrm{plasma}\simeq 9 \ \mathrm{kHz}\left(\frac{n_e}{\mathrm{cm}^{-3}}\right)^{1/2},
    \label{eqn:plasma_freq}
    \end{equation}
    where $n_e$ is the electron density. While the Voyager-2 flyby of the Uranian system was not close enough to measure the electron density near its icy moons, we can bound the cutoff frequency by scaling electron density and surface gravity to other icy moons such as Europa. Using the upper bounds on peak plasma density derived in \S\ref{sec:noise} (see Figure~\ref{fig:plasma_density_UL}) and plugging them into Equation~\ref{eqn:plasma_freq} we obtain an upper limit to the peak plasma frequency. Figure~\ref{fig:plasma_freq_UL} shows the usable frequency band below 900~kHz and above the plasma frequency upper limit (green bars), the band that could potentially be used between the plasma frequency upper limit and the Uranian system's ambient plasma frequency $n_e=1 \ \mathrm{cm^{-3}}$ ($f_\mathrm{plasma}=9 \ \mathrm{kHz}$) (yellow bars) and the frequency band definitely not usable in the Uranian system with $f<9 \ \mathrm{kHz}$ (red bars). In the worst case, the ionospheric cutoff frequency could be as high as $450$~kHz (for Titania), which still allows for a significant part of the UKR spectrum to penetrate into the icy moon. As discussed in \S\ref{sec:noise} these upper limits are aggressively pessimistic. Even with these upper bounds, a significant portion of the spectrum of UKR emissions will penetrate through the ionospheres enabling passive sounding.

\begin{figure}
    \centering\includegraphics[width=0.495\textwidth]{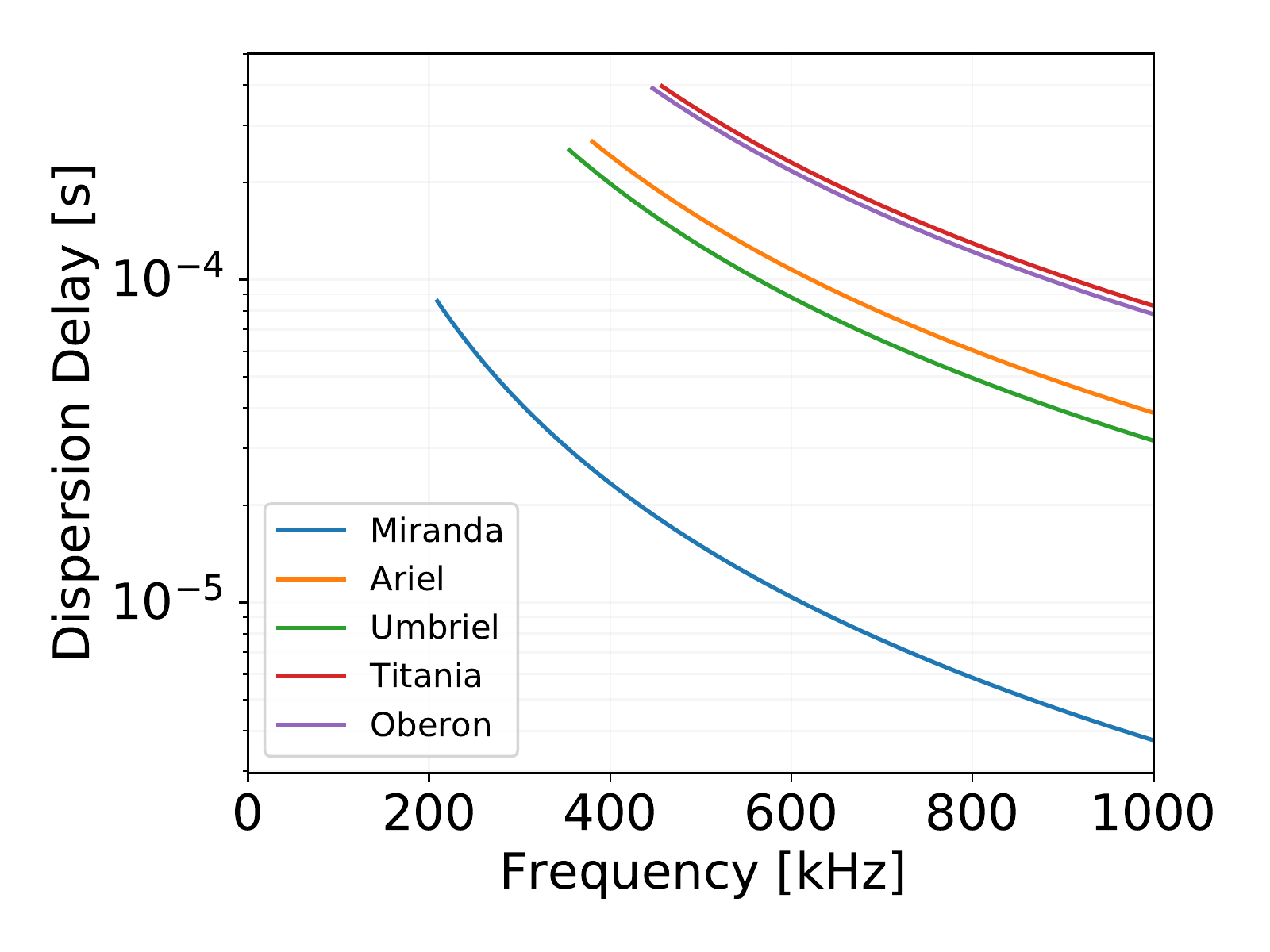}
    \centering\includegraphics[width=0.495\textwidth]{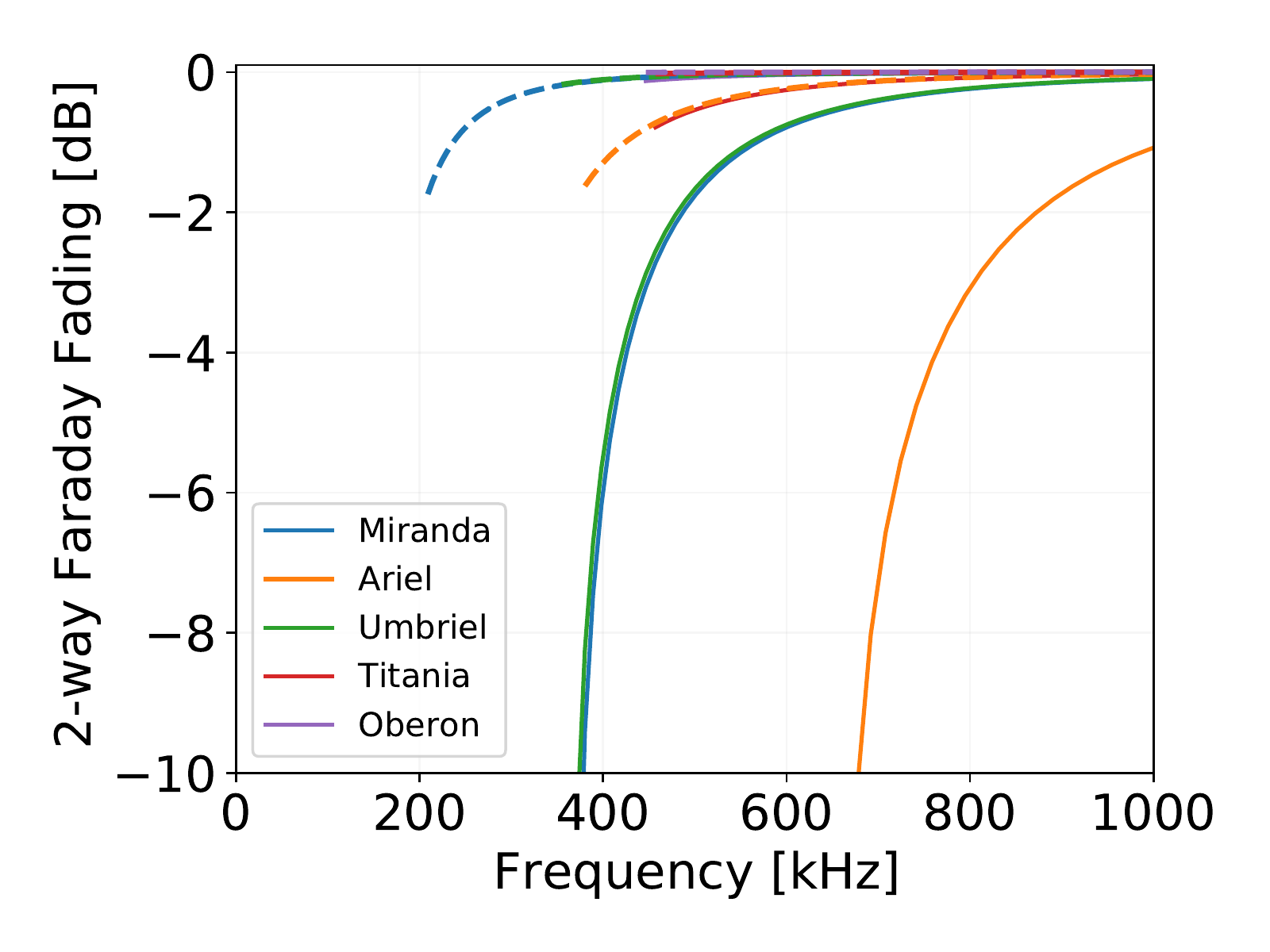}
    \caption{Left: The ionospheric dispersion delay assuming pessimistic ionospheric total electron content (TEC) obtained by scaling Europa's $\mathrm{TEC_{Eu}}\sim4\times10^{15} \  \mathrm{m}^{-2}$ to the square of surface gravity of Uranian icy moons (see text for details). Right: the 2-way Faraday fading vs frequency, as defined in~\cite{Grima_2015}, which provides a loss of signal due to the birefringence of the ionosphere under the influence of the Uranian magnetic field. }
    \label{fig:iono_prop}
    \end{figure}

    We also estimate upper bounds on the impact of the icy moon ionospheres on radio signal propagation (Figure~\ref{fig:iono_prop}). Following~\citeA{Grima_2015} we estimate the ionospheric phase delay due to dispersion for 2-way propagation according to 
    \begin{equation}
        \Delta T_\mathrm{2way} = \frac{2.69\times10^{-7}}{f^2}TEC.
    \end{equation}
    This equation is valid for frequencies above the plasma frequnecy and the gyrofrequency $f_g = 2.8\times10^{10} B$, which is below 10~kHz for the Uranian icy moons. The dispersion delay, including only frequencies above the plasma frequency, is shown in the left panel of Figure~\ref{fig:iono_prop}. The dispersion allows for the use of 10~kHz sub-bands (corresponding to a time resolution of $10^{-4}$~s). For the purposes of estimating an upper bound, we use the total electron content (TEC) of Europa integrate up to an altitude of 1000~km $\mathrm{TEC_{Eu}}\simeq4\times10^{15}$~$\mathrm{m^{-2}}$ and scale it with the square of the surface gravity of the Uranian icy moon to obtain $TEC_{M}\simeq TEC_{Eu}(g_M/g_{Eu})^2$. One factor of the surface gravity comes from the scaling to the peak electron density and a second one comes from the modification of ionospheric scale height. 
    Note that the surface return signal is bright with predictable delays allowing for dispersion effects to be deconvolved and corrected. This same deconvolution would apply to subsurface return signals. 

    The 2-way Faraday fading, as defined in~\cite{Grima_2015}, provides a measure of the signal loss due to Faraday rotation. The right panel of Figure~\ref{fig:iono_prop} shows expected losses due to this effect, again, assuming pessimistic parameters. The solid lines correspond to the magnetic field being aligned with the direction of propagation while the dashed lines are offset by $80^{\circ}$ from the direction of propagation. 
    

    {\it Goniopolarimetry:} The goniopolarimetric technique enables direction finding of circularly polarized radio waves using the correlation between multiple co-located antennas~\cite{Cecconi_2005}. This technique has been applied successfully to the localization of Saturn's kilometric radio source with the RPWS instrument on Cassini~\cite{Cecconi_2009}. This technique could be applied to localizing not only the UKR but also potentially the reflected signal from an icy moon. The analysis presented in~\citeA{Cecconi_2009} did require fairly high signal-to-noise ratio cuts, but in this proposed passive radar system there would be significant increases in time-bandwidth product integration that could enable application of the technique to reflected signals. The localization and polarization vector of the reflected signal could provide further insight into the moon ice shells, particularly for the case of bistatic reflections. The feasibility of goniopolarimetric localization in the context of passive sounding should be further explored in simulations. 

    {\it Occultations:} Observing the transmitted power through an icy moon during a UKR occultation pass could potentially serve as an additional characterization of the ice shell attenuation profile. The reference levels prior to ingress and following egress would serve as reference power levels. Ray-propagation studies would be needed to investigate the sensitivity to various attenuation profile scenarios covered in this paper, including the effects of a potential subsurface reflecting ocean. Studies of Jovian moon occultations with Galileo~\cite{Cecconi_2021} have shown that these measurements can be applied to constraining the source location. While this could be accomplished with goniopolarimetry, as mentioned above, occultations may provide additional constraints on the attenuation profile of the ice shells by measuring the transmission of the UKR. The combination of UKR occultations and goniopolarimetric localization with transmission through the ice could prove a powerful technique, although feasibility needs to be demonstrated via detailed simulations. 
    
    
\section{Conclusions}
    
This initial feasibility assessment of passively sounding Uranian icy moon cryospheres using Uranian Kilometric Radio emissions is promising. We have reached this conclusion after evaluating the source properties, receiver model, target properties, and a range of possible physical models of the Uranian icy moon cryospheres. 

The flux density of the UKR source in the vicinity of Uranian icy moons is orders of magnitude higher than the background identified, meaning the performance for passive sounding is limited only by the integration time-bandwidth product. The source extent is sufficiently compact for a passive sounder to maintain coherence  at reasonable flyby altitudes. The beam pattern and source extent make the UKR source availability predictable with values of at least $\sim55\%$ and possibly as high as $\sim87\%$ if the beam pattern is wider than what was available to the Voyager-2 flyby of Uranus. 

The receiver used for this study is modeled after the RPWS instrument on NASA's Cassini mission but with a modified back end consisting of a 1-MHz instantaneous bandwidth digitizer and signal processing chain. The measured receiver noise floor is significantly below the UKR flux in the vicinity of Uranian icy moons so that it is not necessary to improve on it for passive sounding. The plasma noise will not significantly impact the frequency band of interest provided that the electron density in the plasma surrounding the receiver is $n_e<2500$~cm$^{-3}$, which is an aggressively pessimistic estimate based on scaling Europa's ionosphere and surface gravity, and Voyager-2 measurements in the Uranian system were $n_e\leq1$~cm$^{-3}$. Galactic noise is also a negligible contributor to background noise at the frequencies of interest ($<1$~MHz).

For 
cold oceans, which challenge magnetic induction techniques, passive sounding can directly probe the ice-ocean interface. We predict that losses due to attenuation and scattering due to porosity will be small. If the oceans are warm such that the attenuation prohibits direct ocean detection, brine expected in the lower ice shell, when the ice temperature exceeds the NH$_3$ eutectic temperature, will still be detectable, allowing constraint of the thermal profile of the ice shell. Under these circumstances this method would complement magnetometic induction techniques by directly measuring the ice shell thickness, thus enhancing the ability to characterize ocean properties.

Given this is an initial estimate, we have identified key modeling refinements needed to further develop this concept. Radio emission simulations and the ionospheric density profile expectations are important to understand the uncertainties and provide more accurate estimates of source availability. Future studies of the goniopolarimetric capabilities and UKR occultation by the ice shells would further enrich the understanding of the Uranian icy moon cryospheres.


\section*{Data Availability Statement}
This work uses publicly available data from a variety of sources. 
Figure~\ref{fig:flux} uses UKR average flux density spectrum is from~\citeA{Zarka_1998} and sky background noise spectral density is from~\citeA{Cane_1979}. 
Figure~\ref{fig:max_altitude} uses icy moon radii and orbital distances from \url{https://ssd.jpl.nasa.gov/sats/phys_par/} and \url{https://ssd.jpl.nasa.gov/sats/ephem/}, respectively, source size is obtained from Figure~3~of~\citeA{Menietti_1990} along with the maximum altitude limit provided in~\citeA{Peters_2022}. 
Figure~\ref{fig:illumination_view_angles} samples points in the Figure~3~of~\citeA{Menietti_1990} along with view angles derived from the same geometric parameters used in Figure~\ref{fig:max_altitude}. 
Figure~\ref{fig:noise} uses UKR fluxes from~\citeA{Zarka_1998} along with Cassini RWPS noise and calibration data from~\citeA{Zarka_2004b}. The Galactic flux curve is obtained from data in~\citeA{Manning_2001}. Electron density and temperature parameters are provided in the text and are based on representative values from~\citeA{Sittler_1987} and~\citeA{Kliore_1997}. Same sources were used in Figure~\ref{fig:plasma_density_UL} scaled to the geometric parameters used for Figure~\ref{fig:max_altitude}. 
The data in Table~\ref{tab:structure} is derived from parameters in~\citeA{Castillo2023}.
The curves in Figure~\ref{fig:attenuation} are derived from the data in~Table~\ref{tab:structure} and equations provided in~\S\ref{sec:target} (see references therein). 
The results of Figure~\ref{fig:snr} combine the results of Figure~\ref{fig:attenuation} and the geometric parameters used for Figure~\ref{fig:max_altitude}.
Figure~\ref{fig:plasma_freq_UL} combines data mentioned above (see Figure~\ref{fig:plasma_density_UL}). 
Figure~\ref{fig:iono_prop} uses data from~\citeA{Kliore_1997} scaled by surface gravity obtained from~\url{https://ssd.jpl.nasa.gov/sats/phys_par/} using Equations from~\citeA{Grima_2015}. 

\acknowledgments
The research was carried out in part at the Jet Propulsion Laboratory, California Institute of Technology, under a contract with the National Aeronautics and Space Administration (80NM0018D0004). \copyright 2023. All rights reserved. 


%
\bibliography{agusample}

\begin{thebibliography}{}

\bibitem [\protect \citeauthoryear {%
Aglyamov%
, Schroeder%
\BCBL {}\ \BBA {} Vance%
}{%
Aglyamov%
\ \protect \BOthers {.}}{%
{\protect \APACyear {2017}}%
}]{%
Aglyamov2017}
\APACinsertmetastar {%
Aglyamov2017}%
\begin{APACrefauthors}%
Aglyamov, Y.%
, Schroeder, D\BPBI M.%
\BCBL {}\ \BBA {} Vance, S\BPBI D.%
\end{APACrefauthors}%
\unskip\
\newblock
\APACrefYearMonthDay{2017}{}{}.
\newblock
{\BBOQ}\APACrefatitle {Bright prospects for radar detection of Europa’s
  ocean} {Bright prospects for radar detection of europa’s ocean}.{\BBCQ}
\newblock
\APACjournalVolNumPages{Icarus}{281}{}{334--337}.
\PrintBackRefs{\CurrentBib}

\bibitem [\protect \citeauthoryear {%
{Arridge}%
\ \BBA {} {Eggington}%
}{%
{Arridge}%
\ \BBA {} {Eggington}%
}{%
{\protect \APACyear {2021}}%
}]{%
Arridge_2021}
\APACinsertmetastar {%
Arridge_2021}%
\begin{APACrefauthors}%
{Arridge}, C\BPBI S.%
\BCBT {}\ \BBA {} {Eggington}, J\BPBI W\BPBI B.%
\end{APACrefauthors}%
\unskip\
\newblock
\APACrefYearMonthDay{2021}{{\APACmonth{10}}}{}.
\newblock
{\BBOQ}\APACrefatitle {{Electromagnetic induction in the icy satellites of
  Uranus}} {{Electromagnetic induction in the icy satellites of
  Uranus}}.{\BBCQ}
\newblock
\APACjournalVolNumPages{Icarus}{367}{}{114562}.
\newblock
\begin{APACrefDOI} \doi{10.1016/j.icarus.2021.114562} \end{APACrefDOI}
\PrintBackRefs{\CurrentBib}

\bibitem [\protect \citeauthoryear {%
Balcerak%
}{%
Balcerak%
}{%
{\protect \APACyear {2012}}%
}]{%
balcerak2012}
\APACinsertmetastar {%
balcerak2012}%
\begin{APACrefauthors}%
Balcerak, E.%
\end{APACrefauthors}%
\unskip\
\newblock
\APACrefYearMonthDay{2012}{}{}.
\newblock
{\BBOQ}\APACrefatitle {Uranus's auroras observed from Hubble Space Telescope}
  {Uranus's auroras observed from hubble space telescope}.{\BBCQ}
\newblock
\APACjournalVolNumPages{Eos, Transactions American Geophysical
  Union}{93}{21}{208--208}.
\PrintBackRefs{\CurrentBib}

\bibitem [\protect \citeauthoryear {%
C.~Beddingfield%
, Burr%
\BCBL {}\ \BBA {} Emery%
}{%
C.~Beddingfield%
\ \protect \BOthers {.}}{%
{\protect \APACyear {2015}}%
}]{%
beddingfield2015}
\APACinsertmetastar {%
beddingfield2015}%
\begin{APACrefauthors}%
Beddingfield, C.%
, Burr, D.%
\BCBL {}\ \BBA {} Emery, J.%
\end{APACrefauthors}%
\unskip\
\newblock
\APACrefYearMonthDay{2015}{}{}.
\newblock
{\BBOQ}\APACrefatitle {Fault geometries on Uranus’ satellite Miranda:
  Implications for internal structure and heat flow} {Fault geometries on
  uranus’ satellite miranda: Implications for internal structure and heat
  flow}.{\BBCQ}
\newblock
\APACjournalVolNumPages{Icarus}{247}{}{35--52}.
\PrintBackRefs{\CurrentBib}

\bibitem [\protect \citeauthoryear {%
C\BPBI B.~Beddingfield%
, Cartwright%
, Leonard%
, Nordheim%
\BCBL {}\ \BBA {} Scipioni%
}{%
C\BPBI B.~Beddingfield%
, Cartwright%
\BCBL {}\ \protect \BOthers {.}}{%
{\protect \APACyear {2022}}%
}]{%
beddingfield2022ariel}
\APACinsertmetastar {%
beddingfield2022ariel}%
\begin{APACrefauthors}%
Beddingfield, C\BPBI B.%
, Cartwright, R\BPBI J.%
, Leonard, E.%
, Nordheim, T.%
\BCBL {}\ \BBA {} Scipioni, F.%
\end{APACrefauthors}%
\unskip\
\newblock
\APACrefYearMonthDay{2022}{}{}.
\newblock
{\BBOQ}\APACrefatitle {Ariel's Elastic Thicknesses and Heat Fluxes} {Ariel's
  elastic thicknesses and heat fluxes}.{\BBCQ}
\newblock
\APACjournalVolNumPages{The Planetary Science Journal}{3}{5}{106}.
\PrintBackRefs{\CurrentBib}

\bibitem [\protect \citeauthoryear {%
C\BPBI B.~Beddingfield%
, Leonard%
, Cartwright%
, Elder%
\BCBL {}\ \BBA {} Nordheim%
}{%
C\BPBI B.~Beddingfield%
, Leonard%
\BCBL {}\ \protect \BOthers {.}}{%
{\protect \APACyear {2022}}%
}]{%
beddingfield2022miranda}
\APACinsertmetastar {%
beddingfield2022miranda}%
\begin{APACrefauthors}%
Beddingfield, C\BPBI B.%
, Leonard, E.%
, Cartwright, R\BPBI J.%
, Elder, C.%
\BCBL {}\ \BBA {} Nordheim, T\BPBI A.%
\end{APACrefauthors}%
\unskip\
\newblock
\APACrefYearMonthDay{2022}{}{}.
\newblock
{\BBOQ}\APACrefatitle {High heat flux near Miranda’s Inverness Corona
  consistent with a geologically recent heating event} {High heat flux near
  miranda’s inverness corona consistent with a geologically recent heating
  event}.{\BBCQ}
\newblock
\APACjournalVolNumPages{The Planetary Science Journal}{3}{7}{174}.
\PrintBackRefs{\CurrentBib}

\bibitem [\protect \citeauthoryear {%
Bierson%
\ \BBA {} Nimmo%
}{%
Bierson%
\ \BBA {} Nimmo%
}{%
{\protect \APACyear {2022}}%
}]{%
bierson2022}
\APACinsertmetastar {%
bierson2022}%
\begin{APACrefauthors}%
Bierson, C\BPBI J.%
\BCBT {}\ \BBA {} Nimmo, F.%
\end{APACrefauthors}%
\unskip\
\newblock
\APACrefYearMonthDay{2022}{}{}.
\newblock
{\BBOQ}\APACrefatitle {A note on the possibility of subsurface oceans on the
  uranian satellites} {A note on the possibility of subsurface oceans on the
  uranian satellites}.{\BBCQ}
\newblock
\APACjournalVolNumPages{Icarus}{373}{}{114776}.
\PrintBackRefs{\CurrentBib}

\bibitem [\protect \citeauthoryear {%
{Cane}%
}{%
{Cane}%
}{%
{\protect \APACyear {1979}}%
}]{%
Cane_1979}
\APACinsertmetastar {%
Cane_1979}%
\begin{APACrefauthors}%
{Cane}, H\BPBI V.%
\end{APACrefauthors}%
\unskip\
\newblock
\APACrefYearMonthDay{1979}{{\APACmonth{11}}}{}.
\newblock
{\BBOQ}\APACrefatitle {{Spectra of the non-thermal radio radiation from the
  galactic polar regions.}} {{Spectra of the non-thermal radio radiation from
  the galactic polar regions.}}{\BBCQ}
\newblock
\APACjournalVolNumPages{Monthly Notices of the Royal Astronomical
  Society}{189}{}{465-478}.
\newblock
\begin{APACrefDOI} \doi{10.1093/mnras/189.3.465} \end{APACrefDOI}
\PrintBackRefs{\CurrentBib}

\bibitem [\protect \citeauthoryear {%
Carrer%
, Schroeder%
, Romero-Wolf%
, Ries%
\BCBL {}\ \BBA {} Bruzzone%
}{%
Carrer%
\ \protect \BOthers {.}}{%
{\protect \APACyear {2021}}%
}]{%
Carrer_2021}
\APACinsertmetastar {%
Carrer_2021}%
\begin{APACrefauthors}%
Carrer, L.%
, Schroeder, D\BPBI M.%
, Romero-Wolf, A.%
, Ries, P\BPBI A.%
\BCBL {}\ \BBA {} Bruzzone, L.%
\end{APACrefauthors}%
\unskip\
\newblock
\APACrefYearMonthDay{2021}{}{}.
\newblock
{\BBOQ}\APACrefatitle {Analysis of Temporal and Structural Characteristics of
  Jovian Radio Emissions for Passive Radar Sounding of Jupiter’s Icy Moons}
  {Analysis of temporal and structural characteristics of jovian radio
  emissions for passive radar sounding of jupiter’s icy moons}.{\BBCQ}
\newblock
\APACjournalVolNumPages{IEEE Transactions on Geoscience and Remote
  Sensing}{59}{5}{3857-3874}.
\newblock
\begin{APACrefDOI} \doi{10.1109/TGRS.2020.3023249} \end{APACrefDOI}
\PrintBackRefs{\CurrentBib}

\bibitem [\protect \citeauthoryear {%
Cartwright%
\ \protect \BOthers {.}}{%
Cartwright%
\ \protect \BOthers {.}}{%
{\protect \APACyear {2020}}%
}]{%
cartwright2020}
\APACinsertmetastar {%
cartwright2020}%
\begin{APACrefauthors}%
Cartwright, R\BPBI J.%
, Beddingfield, C\BPBI B.%
, Nordheim, T\BPBI A.%
, Roser, J.%
, Grundy, W\BPBI M.%
, Hand, K\BPBI P.%
\BDBL {}Scipioni, F.%
\end{APACrefauthors}%
\unskip\
\newblock
\APACrefYearMonthDay{2020}{}{}.
\newblock
{\BBOQ}\APACrefatitle {Evidence for ammonia-bearing species on the Uranian
  satellite Ariel supports recent geologic activity} {Evidence for
  ammonia-bearing species on the uranian satellite ariel supports recent
  geologic activity}.{\BBCQ}
\newblock
\APACjournalVolNumPages{The Astrophysical Journal Letters}{898}{1}{L22}.
\PrintBackRefs{\CurrentBib}

\bibitem [\protect \citeauthoryear {%
Cartwright%
\ \protect \BOthers {.}}{%
Cartwright%
\ \protect \BOthers {.}}{%
{\protect \APACyear {2023}}%
}]{%
cartwright2023}
\APACinsertmetastar {%
cartwright2023}%
\begin{APACrefauthors}%
Cartwright, R\BPBI J.%
, DeColibus, D\BPBI R.%
, Castillo-Rogez, J\BPBI C.%
, Beddingfield, C\BPBI B.%
, Grundy, W\BPBI M.%
\BCBL {}\ \BBA {} Nordheim, T\BPBI A.%
\end{APACrefauthors}%
\unskip\
\newblock
\APACrefYearMonthDay{2023}{}{}.
\newblock
{\BBOQ}\APACrefatitle {Evidence for Nitrogen-bearing Species on Umbriel:
  Sourced from a Subsurface Ocean, Undifferentiated Crust, or Impactors?}
  {Evidence for nitrogen-bearing species on umbriel: Sourced from a subsurface
  ocean, undifferentiated crust, or impactors?}{\BBCQ}
\newblock
\APACjournalVolNumPages{The Planetary Science Journal}{4}{3}{42}.
\PrintBackRefs{\CurrentBib}

\bibitem [\protect \citeauthoryear {%
Castillo-Rogez%
\ \protect \BOthers {.}}{%
Castillo-Rogez%
\ \protect \BOthers {.}}{%
{\protect \APACyear {2023}}%
}]{%
Castillo2023}
\APACinsertmetastar {%
Castillo2023}%
\begin{APACrefauthors}%
Castillo-Rogez, J.%
, Weiss, B.%
, Beddingfield, C.%
, Biersteker, J.%
, Cartwright, R.%
, Goode, A.%
\BDBL {}Neveu, M.%
\end{APACrefauthors}%
\unskip\
\newblock
\APACrefYearMonthDay{2023}{}{}.
\newblock
{\BBOQ}\APACrefatitle {Compositions and Interior Structures of the Large Moons
  of Uranus and Implications for Future Spacecraft Observations} {Compositions
  and interior structures of the large moons of uranus and implications for
  future spacecraft observations}.{\BBCQ}
\newblock
\APACjournalVolNumPages{Journal of Geophysical Research:
  Planets}{128}{1}{e2022JE007432}.
\newblock
\begin{APACrefURL}
  \url{https://agupubs.onlinelibrary.wiley.com/doi/abs/10.1029/2022JE007432}
  \end{APACrefURL}
\newblock
\APACrefnote{e2022JE007432 2022JE007432}
\newblock
\begin{APACrefDOI} \doi{https://doi.org/10.1029/2022JE007432} \end{APACrefDOI}
\PrintBackRefs{\CurrentBib}

\bibitem [\protect \citeauthoryear {%
{Cecconi}%
\ \protect \BOthers {.}}{%
{Cecconi}%
\ \protect \BOthers {.}}{%
{\protect \APACyear {2009}}%
}]{%
Cecconi_2009}
\APACinsertmetastar {%
Cecconi_2009}%
\begin{APACrefauthors}%
{Cecconi}, B.%
, {Lamy}, L.%
, {Zarka}, P.%
, {Prang{\'e}}, R.%
, {Kurth}, W\BPBI S.%
\BCBL {}\ \BBA {} {Louarn}, P.%
\end{APACrefauthors}%
\unskip\
\newblock
\APACrefYearMonthDay{2009}{{\APACmonth{03}}}{}.
\newblock
{\BBOQ}\APACrefatitle {{Goniopolarimetric study of the revolution 29 perikrone
  using the Cassini Radio and Plasma Wave Science instrument high-frequency
  radio receiver}} {{Goniopolarimetric study of the revolution 29 perikrone
  using the Cassini Radio and Plasma Wave Science instrument high-frequency
  radio receiver}}.{\BBCQ}
\newblock
\APACjournalVolNumPages{Journal of Geophysical Research (Space
  Physics)}{114}{A3}{A03215}.
\newblock
\begin{APACrefDOI} \doi{10.1029/2008JA013830} \end{APACrefDOI}
\PrintBackRefs{\CurrentBib}

\bibitem [\protect \citeauthoryear {%
{Cecconi}%
, {Louis}%
, {Mu{\~n}oz Crego}%
\BCBL {}\ \BBA {} {Vallat}%
}{%
{Cecconi}%
\ \protect \BOthers {.}}{%
{\protect \APACyear {2021}}%
}]{%
Cecconi_2021}
\APACinsertmetastar {%
Cecconi_2021}%
\begin{APACrefauthors}%
{Cecconi}, B.%
, {Louis}, C\BPBI K.%
, {Mu{\~n}oz Crego}, C.%
\BCBL {}\ \BBA {} {Vallat}, C.%
\end{APACrefauthors}%
\unskip\
\newblock
\APACrefYearMonthDay{2021}{{\APACmonth{12}}}{}.
\newblock
{\BBOQ}\APACrefatitle {{Jovian auroral radio source occultation modelling and
  application to the JUICE science mission planning}} {{Jovian auroral radio
  source occultation modelling and application to the JUICE science mission
  planning}}.{\BBCQ}
\newblock
\APACjournalVolNumPages{Planetary and Space Science}{209}{}{105344}.
\newblock
\begin{APACrefDOI} \doi{10.1016/j.pss.2021.105344} \end{APACrefDOI}
\PrintBackRefs{\CurrentBib}

\bibitem [\protect \citeauthoryear {%
{Cecconi}%
\ \BBA {} {Zarka}%
}{%
{Cecconi}%
\ \BBA {} {Zarka}%
}{%
{\protect \APACyear {2005}}%
}]{%
Cecconi_2005}
\APACinsertmetastar {%
Cecconi_2005}%
\begin{APACrefauthors}%
{Cecconi}, B.%
\BCBT {}\ \BBA {} {Zarka}, P.%
\end{APACrefauthors}%
\unskip\
\newblock
\APACrefYearMonthDay{2005}{{\APACmonth{05}}}{}.
\newblock
{\BBOQ}\APACrefatitle {{Direction finding and antenna calibration through
  analytical inversion of radio measurements performed using a system of two or
  three electric dipole antennas on a three-axis stabilized spacecraft}}
  {{Direction finding and antenna calibration through analytical inversion of
  radio measurements performed using a system of two or three electric dipole
  antennas on a three-axis stabilized spacecraft}}.{\BBCQ}
\newblock
\APACjournalVolNumPages{Radio Science}{40}{3}{RS3003}.
\newblock
\begin{APACrefDOI} \doi{10.1029/2004RS003070} \end{APACrefDOI}
\PrintBackRefs{\CurrentBib}

\bibitem [\protect \citeauthoryear {%
Cochrane%
\ \protect \BOthers {.}}{%
Cochrane%
\ \protect \BOthers {.}}{%
{\protect \APACyear {2021}}%
}]{%
Cochrane2021}
\APACinsertmetastar {%
Cochrane2021}%
\begin{APACrefauthors}%
Cochrane, C\BPBI J.%
, Vance, S\BPBI D.%
, Nordheim, T\BPBI A.%
, Styczinski, M\BPBI J.%
, Masters, A.%
\BCBL {}\ \BBA {} Regoli, L\BPBI H.%
\end{APACrefauthors}%
\unskip\
\newblock
\APACrefYearMonthDay{2021}{}{}.
\newblock
{\BBOQ}\APACrefatitle {In Search of Subsurface Oceans Within the Uranian Moons}
  {In search of subsurface oceans within the uranian moons}.{\BBCQ}
\newblock
\APACjournalVolNumPages{Journal of Geophysical Research:
  Planets}{126}{12}{e2021JE006956}.
\newblock
\begin{APACrefDOI} \doi{https://doi.org/10.1029/2021JE006956} \end{APACrefDOI}
\PrintBackRefs{\CurrentBib}

\bibitem [\protect \citeauthoryear {%
{Cochrane}%
\ \protect \BOthers {.}}{%
{Cochrane}%
\ \protect \BOthers {.}}{%
{\protect \APACyear {2021}}%
}]{%
Cochrane_2021}
\APACinsertmetastar {%
Cochrane_2021}%
\begin{APACrefauthors}%
{Cochrane}, C\BPBI J.%
, {Vance}, S\BPBI D.%
, {Nordheim}, T\BPBI A.%
, {Styczinski}, M\BPBI J.%
, {Masters}, A.%
\BCBL {}\ \BBA {} {Regoli}, L\BPBI H.%
\end{APACrefauthors}%
\unskip\
\newblock
\APACrefYearMonthDay{2021}{{\APACmonth{12}}}{}.
\newblock
{\BBOQ}\APACrefatitle {{In Search of Subsurface Oceans Within the Uranian
  Moons}} {{In Search of Subsurface Oceans Within the Uranian Moons}}.{\BBCQ}
\newblock
\APACjournalVolNumPages{Journal of Geophysical Research
  (Planets)}{126}{12}{e06956}.
\newblock
\begin{APACrefDOI} \doi{10.1029/2021JE006956} \end{APACrefDOI}
\PrintBackRefs{\CurrentBib}

\bibitem [\protect \citeauthoryear {%
Culha%
, Schroeder%
, Jordan%
\BCBL {}\ \BBA {} Haynes%
}{%
Culha%
\ \protect \BOthers {.}}{%
{\protect \APACyear {2020}}%
}]{%
Culha2020}
\APACinsertmetastar {%
Culha2020}%
\begin{APACrefauthors}%
Culha, C.%
, Schroeder, D\BPBI M.%
, Jordan, T\BPBI M.%
\BCBL {}\ \BBA {} Haynes, M\BPBI S.%
\end{APACrefauthors}%
\unskip\
\newblock
\APACrefYearMonthDay{2020}{}{}.
\newblock
{\BBOQ}\APACrefatitle {Assessing the detectability of Europa’s eutectic zone
  using radar sounding} {Assessing the detectability of europa’s eutectic
  zone using radar sounding}.{\BBCQ}
\newblock
\APACjournalVolNumPages{Icarus}{339}{}{113578}.
\PrintBackRefs{\CurrentBib}

\bibitem [\protect \citeauthoryear {%
Eluszkiewicz%
}{%
Eluszkiewicz%
}{%
{\protect \APACyear {2004}}%
}]{%
Eluszkiewicz2004}
\APACinsertmetastar {%
Eluszkiewicz2004}%
\begin{APACrefauthors}%
Eluszkiewicz, J.%
\end{APACrefauthors}%
\unskip\
\newblock
\APACrefYearMonthDay{2004}{}{}.
\newblock
{\BBOQ}\APACrefatitle {Dim prospects for radar detection of Europa's ocean}
  {Dim prospects for radar detection of europa's ocean}.{\BBCQ}
\newblock
\APACjournalVolNumPages{Icarus}{170}{1}{234-236}.
\newblock
\begin{APACrefURL}
  \url{https://www.sciencedirect.com/science/article/pii/S0019103504000788}
  \end{APACrefURL}
\newblock
\begin{APACrefDOI} \doi{https://doi.org/10.1016/j.icarus.2004.02.011}
  \end{APACrefDOI}
\PrintBackRefs{\CurrentBib}

\bibitem [\protect \citeauthoryear {%
FUJINO%
}{%
FUJINO%
}{%
{\protect \APACyear {1967}}%
}]{%
fujino1967}
\APACinsertmetastar {%
fujino1967}%
\begin{APACrefauthors}%
FUJINO, K.%
\end{APACrefauthors}%
\unskip\
\newblock
\APACrefYearMonthDay{1967}{}{}.
\newblock
{\BBOQ}\APACrefatitle {Electrical properties of sea ice} {Electrical properties
  of sea ice}.{\BBCQ}
\newblock
\APACjournalVolNumPages{Physics of Snow and Ice: proceedings}{1}{1}{633--648}.
\PrintBackRefs{\CurrentBib}

\bibitem [\protect \citeauthoryear {%
Grasset%
\ \protect \BOthers {.}}{%
Grasset%
\ \protect \BOthers {.}}{%
{\protect \APACyear {2013}}%
}]{%
grasset2013}
\APACinsertmetastar {%
grasset2013}%
\begin{APACrefauthors}%
Grasset, O.%
, Dougherty, M.%
, Coustenis, A.%
, Bunce, E.%
, Erd, C.%
, Titov, D.%
\BDBL {}Fletcher, L.%
\end{APACrefauthors}%
\unskip\
\newblock
\APACrefYearMonthDay{2013}{}{}.
\newblock
{\BBOQ}\APACrefatitle {JUpiter ICy moons Explorer (JUICE): An ESA mission to
  orbit Ganymede and to characterise the Jupiter system} {Jupiter icy moons
  explorer (juice): An esa mission to orbit ganymede and to characterise the
  jupiter system}.{\BBCQ}
\newblock
\APACjournalVolNumPages{Planetary and Space Science}{78}{}{1--21}.
\PrintBackRefs{\CurrentBib}

\bibitem [\protect \citeauthoryear {%
{Grima}%
, {Blankenship}%
\BCBL {}\ \BBA {} {Schroeder}%
}{%
{Grima}%
\ \protect \BOthers {.}}{%
{\protect \APACyear {2015}}%
}]{%
Grima_2015}
\APACinsertmetastar {%
Grima_2015}%
\begin{APACrefauthors}%
{Grima}, C.%
, {Blankenship}, D\BPBI D.%
\BCBL {}\ \BBA {} {Schroeder}, D\BPBI M.%
\end{APACrefauthors}%
\unskip\
\newblock
\APACrefYearMonthDay{2015}{{\APACmonth{11}}}{}.
\newblock
{\BBOQ}\APACrefatitle {{Radar signal propagation through the ionosphere of
  Europa}} {{Radar signal propagation through the ionosphere of
  Europa}}.{\BBCQ}
\newblock
\APACjournalVolNumPages{Planetary and Space Science}{117}{}{421-428}.
\newblock
\begin{APACrefDOI} \doi{10.1016/j.pss.2015.08.017} \end{APACrefDOI}
\PrintBackRefs{\CurrentBib}

\bibitem [\protect \citeauthoryear {%
Gross%
, Wong%
\BCBL {}\ \BBA {} Humes%
}{%
Gross%
\ \protect \BOthers {.}}{%
{\protect \APACyear {1977}}%
}]{%
gross1977concentration}
\APACinsertmetastar {%
gross1977concentration}%
\begin{APACrefauthors}%
Gross, G\BPBI W.%
, Wong, P\BPBI M.%
\BCBL {}\ \BBA {} Humes, K.%
\end{APACrefauthors}%
\unskip\
\newblock
\APACrefYearMonthDay{1977}{}{}.
\newblock
{\BBOQ}\APACrefatitle {Concentration dependent solute redistribution at the
  ice--water phase boundary. III. Spontaneous convection. Chloride solutions}
  {Concentration dependent solute redistribution at the ice--water phase
  boundary. iii. spontaneous convection. chloride solutions}.{\BBCQ}
\newblock
\APACjournalVolNumPages{The Journal of chemical physics}{67}{11}{5264--5274}.
\PrintBackRefs{\CurrentBib}

\bibitem [\protect \citeauthoryear {%
Gulkis%
\ \BBA {} Carr%
}{%
Gulkis%
\ \BBA {} Carr%
}{%
{\protect \APACyear {1987}}%
}]{%
gulkis1987}
\APACinsertmetastar {%
gulkis1987}%
\begin{APACrefauthors}%
Gulkis, S.%
\BCBT {}\ \BBA {} Carr, T\BPBI D.%
\end{APACrefauthors}%
\unskip\
\newblock
\APACrefYearMonthDay{1987}{}{}.
\newblock
{\BBOQ}\APACrefatitle {The main source of radio emission from the magnetosphere
  of Uranus} {The main source of radio emission from the magnetosphere of
  uranus}.{\BBCQ}
\newblock
\APACjournalVolNumPages{Journal of Geophysical Research: Space
  Physics}{92}{A13}{15159--15168}.
\PrintBackRefs{\CurrentBib}

\bibitem [\protect \citeauthoryear {%
{Gurnett}%
\ \protect \BOthers {.}}{%
{Gurnett}%
\ \protect \BOthers {.}}{%
{\protect \APACyear {2004}}%
}]{%
Gurnett_2004}
\APACinsertmetastar {%
Gurnett_2004}%
\begin{APACrefauthors}%
{Gurnett}, D\BPBI A.%
, {Kurth}, W\BPBI S.%
, {Kirchner}, D\BPBI L.%
, {Hospodarsky}, G\BPBI B.%
, {Averkamp}, T\BPBI F.%
, {Zarka}, P.%
\BDBL {}{Pedersen}, A.%
\end{APACrefauthors}%
\unskip\
\newblock
\APACrefYearMonthDay{2004}{{\APACmonth{09}}}{}.
\newblock
{\BBOQ}\APACrefatitle {{The Cassini Radio and Plasma Wave Investigation}} {{The
  Cassini Radio and Plasma Wave Investigation}}.{\BBCQ}
\newblock
\APACjournalVolNumPages{Space Science Reviews}{114}{1-4}{395-463}.
\newblock
\begin{APACrefDOI} \doi{10.1007/s11214-004-1434-0} \end{APACrefDOI}
\PrintBackRefs{\CurrentBib}

\bibitem [\protect \citeauthoryear {%
Hendrix%
\ \protect \BOthers {.}}{%
Hendrix%
\ \protect \BOthers {.}}{%
{\protect \APACyear {2019}}%
}]{%
hendrix2019}
\APACinsertmetastar {%
hendrix2019}%
\begin{APACrefauthors}%
Hendrix, A\BPBI R.%
, Hurford, T\BPBI A.%
, Barge, L\BPBI M.%
, Bland, M\BPBI T.%
, Bowman, J\BPBI S.%
, Brinckerhoff, W.%
\BDBL {}Collins, G\BPBI C.%
\end{APACrefauthors}%
\unskip\
\newblock
\APACrefYearMonthDay{2019}{}{}.
\newblock
{\BBOQ}\APACrefatitle {The NASA roadmap to ocean worlds} {The nasa roadmap to
  ocean worlds}.{\BBCQ}
\newblock
\APACjournalVolNumPages{Astrobiology}{19}{1}{1--27}.
\PrintBackRefs{\CurrentBib}

\bibitem [\protect \citeauthoryear {%
Hussmann%
, Sohl%
\BCBL {}\ \BBA {} Spohn%
}{%
Hussmann%
\ \protect \BOthers {.}}{%
{\protect \APACyear {2006}}%
}]{%
Hussmann2006}
\APACinsertmetastar {%
Hussmann2006}%
\begin{APACrefauthors}%
Hussmann, H.%
, Sohl, F.%
\BCBL {}\ \BBA {} Spohn, T.%
\end{APACrefauthors}%
\unskip\
\newblock
\APACrefYearMonthDay{2006}{}{}.
\newblock
{\BBOQ}\APACrefatitle {Subsurface oceans and deep interiors of medium-sized
  outer planet satellites and large trans-neptunian objects} {Subsurface oceans
  and deep interiors of medium-sized outer planet satellites and large
  trans-neptunian objects}.{\BBCQ}
\newblock
\APACjournalVolNumPages{Icarus}{185}{1}{258--273}.
\PrintBackRefs{\CurrentBib}

\bibitem [\protect \citeauthoryear {%
Kalousová%
, Schroeder%
\BCBL {}\ \BBA {} Soderlund%
}{%
Kalousová%
\ \protect \BOthers {.}}{%
{\protect \APACyear {2017}}%
}]{%
Kalousova2017}
\APACinsertmetastar {%
Kalousova2017}%
\begin{APACrefauthors}%
Kalousová, K.%
, Schroeder, D\BPBI M.%
\BCBL {}\ \BBA {} Soderlund, K\BPBI M.%
\end{APACrefauthors}%
\unskip\
\newblock
\APACrefYearMonthDay{2017}{}{}.
\newblock
{\BBOQ}\APACrefatitle {Radar attenuation in Europa's ice shell: Obstacles and
  opportunities for constraining the shell thickness and its thermal structure}
  {Radar attenuation in europa's ice shell: Obstacles and opportunities for
  constraining the shell thickness and its thermal structure}.{\BBCQ}
\newblock
\APACjournalVolNumPages{Journal of Geophysical Research:
  Planets}{122}{3}{524-545}.
\newblock
\begin{APACrefDOI} \doi{https://doi.org/10.1002/2016JE005110} \end{APACrefDOI}
\PrintBackRefs{\CurrentBib}

\bibitem [\protect \citeauthoryear {%
{Khurana}%
\ \protect \BOthers {.}}{%
{Khurana}%
\ \protect \BOthers {.}}{%
{\protect \APACyear {2011}}%
}]{%
Khurana_2011}
\APACinsertmetastar {%
Khurana_2011}%
\begin{APACrefauthors}%
{Khurana}, K\BPBI K.%
, {Jia}, X.%
, {Kivelson}, M\BPBI G.%
, {Nimmo}, F.%
, {Schubert}, G.%
\BCBL {}\ \BBA {} {Russell}, C\BPBI T.%
\end{APACrefauthors}%
\unskip\
\newblock
\APACrefYearMonthDay{2011}{{\APACmonth{06}}}{}.
\newblock
{\BBOQ}\APACrefatitle {{Evidence of a Global Magma Ocean in
  Io{\textquoteright}s Interior}} {{Evidence of a Global Magma Ocean in
  Io{\textquoteright}s Interior}}.{\BBCQ}
\newblock
\APACjournalVolNumPages{Science}{332}{6034}{1186}.
\newblock
\begin{APACrefDOI} \doi{10.1126/science.1201425} \end{APACrefDOI}
\PrintBackRefs{\CurrentBib}

\bibitem [\protect \citeauthoryear {%
{Kirchoff}%
, {Dones}%
, {Singer}%
\BCBL {}\ \BBA {} {Schenk}%
}{%
{Kirchoff}%
\ \protect \BOthers {.}}{%
{\protect \APACyear {2022}}%
}]{%
Kirchoff_2022}
\APACinsertmetastar {%
Kirchoff_2022}%
\begin{APACrefauthors}%
{Kirchoff}, M\BPBI R.%
, {Dones}, L.%
, {Singer}, K.%
\BCBL {}\ \BBA {} {Schenk}, P.%
\end{APACrefauthors}%
\unskip\
\newblock
\APACrefYearMonthDay{2022}{{\APACmonth{02}}}{}.
\newblock
{\BBOQ}\APACrefatitle {{Crater Distributions of Uranus's Mid-sized Satellites
  and Implications for Outer Solar System Bombardment}} {{Crater Distributions
  of Uranus's Mid-sized Satellites and Implications for Outer Solar System
  Bombardment}}.{\BBCQ}
\newblock
\APACjournalVolNumPages{The Planetary Science Journal}{3}{2}{42}.
\newblock
\begin{APACrefDOI} \doi{10.3847/PSJ/ac42d7} \end{APACrefDOI}
\PrintBackRefs{\CurrentBib}

\bibitem [\protect \citeauthoryear {%
{Kivelson}%
\ \protect \BOthers {.}}{%
{Kivelson}%
\ \protect \BOthers {.}}{%
{\protect \APACyear {1999}}%
}]{%
Kivelson_1999}
\APACinsertmetastar {%
Kivelson_1999}%
\begin{APACrefauthors}%
{Kivelson}, M\BPBI G.%
, {Khurana}, K\BPBI K.%
, {Stevenson}, D\BPBI J.%
, {Bennett}, L.%
, {Joy}, S.%
, {Russell}, C\BPBI T.%
\BDBL {}{Polanskey}, C.%
\end{APACrefauthors}%
\unskip\
\newblock
\APACrefYearMonthDay{1999}{{\APACmonth{03}}}{}.
\newblock
{\BBOQ}\APACrefatitle {{Europa and Callisto: Induced or intrinsic fields in a
  periodically varying plasma environment}} {{Europa and Callisto: Induced or
  intrinsic fields in a periodically varying plasma environment}}.{\BBCQ}
\newblock
\APACjournalVolNumPages{Journal of Geophysical Research}{104}{A3}{4609-4626}.
\newblock
\begin{APACrefDOI} \doi{10.1029/1998JA900095} \end{APACrefDOI}
\PrintBackRefs{\CurrentBib}

\bibitem [\protect \citeauthoryear {%
{Kivelson}%
, {Khurana}%
\BCBL {}\ \BBA {} {Volwerk}%
}{%
{Kivelson}%
\ \protect \BOthers {.}}{%
{\protect \APACyear {2002}}%
}]{%
Kivelson_2002}
\APACinsertmetastar {%
Kivelson_2002}%
\begin{APACrefauthors}%
{Kivelson}, M\BPBI G.%
, {Khurana}, K\BPBI K.%
\BCBL {}\ \BBA {} {Volwerk}, M.%
\end{APACrefauthors}%
\unskip\
\newblock
\APACrefYearMonthDay{2002}{{\APACmonth{06}}}{}.
\newblock
{\BBOQ}\APACrefatitle {{The Permanent and Inductive Magnetic Moments of
  Ganymede}} {{The Permanent and Inductive Magnetic Moments of
  Ganymede}}.{\BBCQ}
\newblock
\APACjournalVolNumPages{Icarus}{157}{2}{507-522}.
\newblock
\begin{APACrefDOI} \doi{10.1006/icar.2002.6834} \end{APACrefDOI}
\PrintBackRefs{\CurrentBib}

\bibitem [\protect \citeauthoryear {%
{Kliore}%
, {Hinson}%
, {Flasar}%
, {Nagy}%
\BCBL {}\ \BBA {} {Cravens}%
}{%
{Kliore}%
\ \protect \BOthers {.}}{%
{\protect \APACyear {1997}}%
}]{%
Kliore_1997}
\APACinsertmetastar {%
Kliore_1997}%
\begin{APACrefauthors}%
{Kliore}, A\BPBI J.%
, {Hinson}, D\BPBI P.%
, {Flasar}, F\BPBI M.%
, {Nagy}, A\BPBI F.%
\BCBL {}\ \BBA {} {Cravens}, T\BPBI E.%
\end{APACrefauthors}%
\unskip\
\newblock
\APACrefYearMonthDay{1997}{{\APACmonth{07}}}{}.
\newblock
{\BBOQ}\APACrefatitle {{The ionosphere of Europa from Galileo radio
  occultations}} {{The ionosphere of Europa from Galileo radio
  occultations}}.{\BBCQ}
\newblock
\APACjournalVolNumPages{Science}{277}{}{355-358}.
\newblock
\begin{APACrefDOI} \doi{10.1126/science.277.5324.355} \end{APACrefDOI}
\PrintBackRefs{\CurrentBib}

\bibitem [\protect \citeauthoryear {%
Krupar%
\ \BBA {} Szabo%
}{%
Krupar%
\ \BBA {} Szabo%
}{%
{\protect \APACyear {2018}}%
}]{%
Krupar_2018}
\APACinsertmetastar {%
Krupar_2018}%
\begin{APACrefauthors}%
Krupar, V.%
\BCBT {}\ \BBA {} Szabo, A.%
\end{APACrefauthors}%
\unskip\
\newblock
\APACrefYearMonthDay{2018}{}{}.
\newblock
{\BBOQ}\APACrefatitle {Statistical Survey of Interplanetary Type II and Type
  III Radio Bursts} {Statistical survey of interplanetary type ii and type iii
  radio bursts}.{\BBCQ}
\newblock
\APACjournalVolNumPages{USRI Proceedings}{}{}{}.
\PrintBackRefs{\CurrentBib}

\bibitem [\protect \citeauthoryear {%
{Lamy}%
\ \protect \BOthers {.}}{%
{Lamy}%
\ \protect \BOthers {.}}{%
{\protect \APACyear {2017}}%
}]{%
Lamy_2017}
\APACinsertmetastar {%
Lamy_2017}%
\begin{APACrefauthors}%
{Lamy}, L.%
, {Prang{\'e}}, R.%
, {Hansen}, K\BPBI C.%
, {Tao}, C.%
, {Cowley}, S\BPBI W\BPBI H.%
, {Stallard}, T\BPBI S.%
\BDBL {}{Pogorelov}, N.%
\end{APACrefauthors}%
\unskip\
\newblock
\APACrefYearMonthDay{2017}{{\APACmonth{04}}}{}.
\newblock
{\BBOQ}\APACrefatitle {{The aurorae of Uranus past equinox}} {{The aurorae of
  Uranus past equinox}}.{\BBCQ}
\newblock
\APACjournalVolNumPages{Journal of Geophysical Research (Space
  Physics)}{122}{4}{3997-4008}.
\newblock
\begin{APACrefDOI} \doi{10.1002/2017JA023918} \end{APACrefDOI}
\PrintBackRefs{\CurrentBib}

\bibitem [\protect \citeauthoryear {%
{Louis}%
\ \protect \BOthers {.}}{%
{Louis}%
\ \protect \BOthers {.}}{%
{\protect \APACyear {2019}}%
}]{%
Louis_2019}
\APACinsertmetastar {%
Louis_2019}%
\begin{APACrefauthors}%
{Louis}, C\BPBI K.%
, {Hess}, S\BPBI L\BPBI G.%
, {Cecconi}, B.%
, {Zarka}, P.%
, {Lamy}, L.%
, {Aicardi}, S.%
\BCBL {}\ \BBA {} {Loh}, A.%
\end{APACrefauthors}%
\unskip\
\newblock
\APACrefYearMonthDay{2019}{{\APACmonth{07}}}{}.
\newblock
{\BBOQ}\APACrefatitle {{ExPRES: an Exoplanetary and Planetary Radio Emissions
  Simulator}} {{ExPRES: an Exoplanetary and Planetary Radio Emissions
  Simulator}}.{\BBCQ}
\newblock
\APACjournalVolNumPages{Astronomy \& Astrophysics}{627}{}{A30}.
\newblock
\begin{APACrefDOI} \doi{10.1051/0004-6361/201935161} \end{APACrefDOI}
\PrintBackRefs{\CurrentBib}

\bibitem [\protect \citeauthoryear {%
MacGregor%
\ \protect \BOthers {.}}{%
MacGregor%
\ \protect \BOthers {.}}{%
{\protect \APACyear {2015}}%
}]{%
MacGregor2015}
\APACinsertmetastar {%
MacGregor2015}%
\begin{APACrefauthors}%
MacGregor, J\BPBI A.%
, Li, J.%
, Paden, J\BPBI D.%
, Catania, G\BPBI A.%
, Clow, G\BPBI D.%
, Fahnestock, M\BPBI A.%
\BDBL {}Stillman, D\BPBI E.%
\end{APACrefauthors}%
\unskip\
\newblock
\APACrefYearMonthDay{2015}{}{}.
\newblock
{\BBOQ}\APACrefatitle {Radar attenuation and temperature within the Greenland
  Ice Sheet} {Radar attenuation and temperature within the greenland ice
  sheet}.{\BBCQ}
\newblock
\APACjournalVolNumPages{Journal of Geophysical Research: Earth
  Surface}{120}{6}{983-1008}.
\newblock
\begin{APACrefURL}
  \url{https://agupubs.onlinelibrary.wiley.com/doi/abs/10.1002/2014JF003418}
  \end{APACrefURL}
\newblock
\begin{APACrefDOI} \doi{https://doi.org/10.1002/2014JF003418} \end{APACrefDOI}
\PrintBackRefs{\CurrentBib}

\bibitem [\protect \citeauthoryear {%
{Manning}%
\ \BBA {} {Dulk}%
}{%
{Manning}%
\ \BBA {} {Dulk}%
}{%
{\protect \APACyear {2001}}%
}]{%
Manning_2001}
\APACinsertmetastar {%
Manning_2001}%
\begin{APACrefauthors}%
{Manning}, R.%
\BCBT {}\ \BBA {} {Dulk}, G\BPBI A.%
\end{APACrefauthors}%
\unskip\
\newblock
\APACrefYearMonthDay{2001}{{\APACmonth{06}}}{}.
\newblock
{\BBOQ}\APACrefatitle {{The Galactic background radiation from 0.2 to 13.8
  MHz}} {{The Galactic background radiation from 0.2 to 13.8 MHz}}.{\BBCQ}
\newblock
\APACjournalVolNumPages{Astronomy and Astrophysics}{372}{}{663-666}.
\newblock
\begin{APACrefDOI} \doi{10.1051/0004-6361:20010516} \end{APACrefDOI}
\PrintBackRefs{\CurrentBib}

\bibitem [\protect \citeauthoryear {%
{Menietti}%
, {Wong}%
, {Wah}%
\BCBL {}\ \BBA {} {Lin}%
}{%
{Menietti}%
\ \protect \BOthers {.}}{%
{\protect \APACyear {1990}}%
}]{%
Menietti_1990}
\APACinsertmetastar {%
Menietti_1990}%
\begin{APACrefauthors}%
{Menietti}, J\BPBI D.%
, {Wong}, H\BPBI K.%
, {Wah}, D\BPBI A.%
\BCBL {}\ \BBA {} {Lin}, C\BPBI S.%
\end{APACrefauthors}%
\unskip\
\newblock
\APACrefYearMonthDay{1990}{{\APACmonth{01}}}{}.
\newblock
{\BBOQ}\APACrefatitle {{Source region of the smooth high-frequency nightside
  Uranus kilometer radiation: A ray-tracing study}} {{Source region of the
  smooth high-frequency nightside Uranus kilometer radiation: A ray-tracing
  study}}.{\BBCQ}
\newblock
\APACjournalVolNumPages{Journal of Geophysical Research}{95}{A1}{51-60}.
\newblock
\begin{APACrefDOI} \doi{10.1029/JA095iA01p00051} \end{APACrefDOI}
\PrintBackRefs{\CurrentBib}

\bibitem [\protect \citeauthoryear {%
{Meyer-Vernet}%
\ \BBA {} {Perche}%
}{%
{Meyer-Vernet}%
\ \BBA {} {Perche}%
}{%
{\protect \APACyear {1989}}%
}]{%
Meyer-Vernet_1989}
\APACinsertmetastar {%
Meyer-Vernet_1989}%
\begin{APACrefauthors}%
{Meyer-Vernet}, N.%
\BCBT {}\ \BBA {} {Perche}, C.%
\end{APACrefauthors}%
\unskip\
\newblock
\APACrefYearMonthDay{1989}{{\APACmonth{03}}}{}.
\newblock
{\BBOQ}\APACrefatitle {{Tool kit for antennae and thermal noise near the plasma
  frequency}} {{Tool kit for antennae and thermal noise near the plasma
  frequency}}.{\BBCQ}
\newblock
\APACjournalVolNumPages{Journal of Geophysical Research}{94}{}{2405-2415}.
\newblock
\begin{APACrefDOI} \doi{10.1029/JA094iA03p02405} \end{APACrefDOI}
\PrintBackRefs{\CurrentBib}

\bibitem [\protect \citeauthoryear {%
{Ndacyayisenga}%
, {Uwamahoro}%
, {Sasikumar Raja}%
\BCBL {}\ \BBA {} {Monstein}%
}{%
{Ndacyayisenga}%
\ \protect \BOthers {.}}{%
{\protect \APACyear {2021}}%
}]{%
Ndacyayisenga_2021}
\APACinsertmetastar {%
Ndacyayisenga_2021}%
\begin{APACrefauthors}%
{Ndacyayisenga}, T.%
, {Uwamahoro}, J.%
, {Sasikumar Raja}, K.%
\BCBL {}\ \BBA {} {Monstein}, C.%
\end{APACrefauthors}%
\unskip\
\newblock
\APACrefYearMonthDay{2021}{{\APACmonth{02}}}{}.
\newblock
{\BBOQ}\APACrefatitle {{A statistical study of solar radio Type III bursts and
  space weather implication}} {{A statistical study of solar radio Type III
  bursts and space weather implication}}.{\BBCQ}
\newblock
\APACjournalVolNumPages{Advances in Space Research}{67}{4}{1425-1435}.
\newblock
\begin{APACrefDOI} \doi{10.1016/j.asr.2020.11.022} \end{APACrefDOI}
\PrintBackRefs{\CurrentBib}

\bibitem [\protect \citeauthoryear {%
Peters%
, Roberts%
, Nessly%
, Schroeder%
\BCBL {}\ \BBA {} Romero-Wolf%
}{%
Peters%
\ \protect \BOthers {.}}{%
{\protect \APACyear {2022}}%
}]{%
Peters_2022}
\APACinsertmetastar {%
Peters_2022}%
\begin{APACrefauthors}%
Peters, S\BPBI T.%
, Roberts, T\BPBI M.%
, Nessly, K.%
, Schroeder, D\BPBI M.%
\BCBL {}\ \BBA {} Romero-Wolf, A.%
\end{APACrefauthors}%
\unskip\
\newblock
\APACrefYearMonthDay{2022}{}{}.
\newblock
{\BBOQ}\APACrefatitle {Revisiting the Limits of Spatial Coherence for Passive
  Radar Sounding Using Radio-Astronomical Sources} {Revisiting the limits of
  spatial coherence for passive radar sounding using radio-astronomical
  sources}.{\BBCQ}
\newblock
\BIn{} \APACrefbtitle {IGARSS 2022 - 2022 IEEE International Geoscience and
  Remote Sensing Symposium} {Igarss 2022 - 2022 ieee international geoscience
  and remote sensing symposium}\ (\BPG~3880-3883).
\newblock
\begin{APACrefDOI} \doi{10.1109/IGARSS46834.2022.9884673} \end{APACrefDOI}
\PrintBackRefs{\CurrentBib}

\bibitem [\protect \citeauthoryear {%
Peters%
, Schroeder%
, Castelletti%
, Haynes%
\BCBL {}\ \BBA {} Romero-Wolf%
}{%
Peters%
\ \protect \BOthers {.}}{%
{\protect \APACyear {2018}}%
}]{%
Peters_2018}
\APACinsertmetastar {%
Peters_2018}%
\begin{APACrefauthors}%
Peters, S\BPBI T.%
, Schroeder, D\BPBI M.%
, Castelletti, D.%
, Haynes, M.%
\BCBL {}\ \BBA {} Romero-Wolf, A.%
\end{APACrefauthors}%
\unskip\
\newblock
\APACrefYearMonthDay{2018}{}{}.
\newblock
{\BBOQ}\APACrefatitle {In Situ Demonstration of a Passive Radio Sounding
  Approach Using the Sun for Echo Detection} {In situ demonstration of a
  passive radio sounding approach using the sun for echo detection}.{\BBCQ}
\newblock
\APACjournalVolNumPages{IEEE Transactions on Geoscience and Remote
  Sensing}{56}{12}{7338-7349}.
\newblock
\begin{APACrefDOI} \doi{10.1109/TGRS.2018.2850662} \end{APACrefDOI}
\PrintBackRefs{\CurrentBib}

\bibitem [\protect \citeauthoryear {%
{Peters}%
\ \protect \BOthers {.}}{%
{Peters}%
\ \protect \BOthers {.}}{%
{\protect \APACyear {2021}}%
}]{%
Peters_2021b}
\APACinsertmetastar {%
Peters_2021b}%
\begin{APACrefauthors}%
{Peters}, S\BPBI T.%
, {Schroeder}, D\BPBI M.%
, {Chu}, W.%
, {Castelletti}, D.%
, {Haynes}, M\BPBI S.%
, {Christoffersen}, P.%
\BCBL {}\ \BBA {} {Romero-Wolf}, A.%
\end{APACrefauthors}%
\unskip\
\newblock
\APACrefYearMonthDay{2021}{{\APACmonth{07}}}{}.
\newblock
{\BBOQ}\APACrefatitle {{Glaciological Monitoring Using the Sun as a Radio
  Source for Echo Detection}} {{Glaciological Monitoring Using the Sun as a
  Radio Source for Echo Detection}}.{\BBCQ}
\newblock
\APACjournalVolNumPages{Geophysical Research Letters}{48}{14}{e92450}.
\newblock
\begin{APACrefDOI} \doi{10.1029/2021GL092450} \end{APACrefDOI}
\PrintBackRefs{\CurrentBib}

\bibitem [\protect \citeauthoryear {%
Peters%
, Schroeder%
, Haynes%
, Castelletti%
\BCBL {}\ \BBA {} Romero-Wolf%
}{%
Peters%
\ \protect \BOthers {.}}{%
{\protect \APACyear {2021}}%
}]{%
Peters_2021a}
\APACinsertmetastar {%
Peters_2021a}%
\begin{APACrefauthors}%
Peters, S\BPBI T.%
, Schroeder, D\BPBI M.%
, Haynes, M\BPBI S.%
, Castelletti, D.%
\BCBL {}\ \BBA {} Romero-Wolf, A.%
\end{APACrefauthors}%
\unskip\
\newblock
\APACrefYearMonthDay{2021}{}{}.
\newblock
{\BBOQ}\APACrefatitle {Passive Synthetic Aperture Radar Imaging Using
  Radio-Astronomical Sources} {Passive synthetic aperture radar imaging using
  radio-astronomical sources}.{\BBCQ}
\newblock
\APACjournalVolNumPages{IEEE Transactions on Geoscience and Remote
  Sensing}{59}{11}{9144-9159}.
\newblock
\begin{APACrefDOI} \doi{10.1109/TGRS.2021.3050429} \end{APACrefDOI}
\PrintBackRefs{\CurrentBib}

\bibitem [\protect \citeauthoryear {%
Phillips%
\ \BBA {} Pappalardo%
}{%
Phillips%
\ \BBA {} Pappalardo%
}{%
{\protect \APACyear {2014}}%
}]{%
phillips2014}
\APACinsertmetastar {%
phillips2014}%
\begin{APACrefauthors}%
Phillips, C\BPBI B.%
\BCBT {}\ \BBA {} Pappalardo, R\BPBI T.%
\end{APACrefauthors}%
\unskip\
\newblock
\APACrefYearMonthDay{2014}{}{}.
\newblock
{\BBOQ}\APACrefatitle {Europa Clipper mission concept: Exploring Jupiter's
  ocean moon} {Europa clipper mission concept: Exploring jupiter's ocean
  moon}.{\BBCQ}
\newblock
\APACjournalVolNumPages{Eos, Transactions American Geophysical
  Union}{95}{20}{165--167}.
\PrintBackRefs{\CurrentBib}

\bibitem [\protect \citeauthoryear {%
Podolak%
, Hubbard%
\BCBL {}\ \BBA {} Stevenson%
}{%
Podolak%
\ \protect \BOthers {.}}{%
{\protect \APACyear {1991}}%
}]{%
podolak1991}
\APACinsertmetastar {%
podolak1991}%
\begin{APACrefauthors}%
Podolak, M.%
, Hubbard, W.%
\BCBL {}\ \BBA {} Stevenson, D.%
\end{APACrefauthors}%
\unskip\
\newblock
\APACrefYearMonthDay{1991}{}{}.
\newblock
{\BBOQ}\APACrefatitle {Models of Uranus’ interior and magnetic field} {Models
  of uranus’ interior and magnetic field}.{\BBCQ}
\newblock
\APACjournalVolNumPages{Uranus}{1}{}{29--61}.
\PrintBackRefs{\CurrentBib}

\bibitem [\protect \citeauthoryear {%
Roberts%
\ \protect \BOthers {.}}{%
Roberts%
\ \protect \BOthers {.}}{%
{\protect \APACyear {2022}}%
}]{%
Roberts_2022}
\APACinsertmetastar {%
Roberts_2022}%
\begin{APACrefauthors}%
Roberts, T\BPBI M.%
, Romero-Wolf, A.%
, Bruzzone, L.%
, Carrer, L.%
, Peters, S.%
\BCBL {}\ \BBA {} Schroeder, D\BPBI M.%
\end{APACrefauthors}%
\unskip\
\newblock
\APACrefYearMonthDay{2022}{}{}.
\newblock
{\BBOQ}\APACrefatitle {Conditioning Jovian Burst Signals for Passive Sounding
  Applications} {Conditioning jovian burst signals for passive sounding
  applications}.{\BBCQ}
\newblock
\APACjournalVolNumPages{IEEE Transactions on Geoscience and Remote
  Sensing}{60}{}{1-14}.
\newblock
\begin{APACrefDOI} \doi{10.1109/TGRS.2021.3109106} \end{APACrefDOI}
\PrintBackRefs{\CurrentBib}

\bibitem [\protect \citeauthoryear {%
{Romero-Wolf}%
\ \protect \BOthers {.}}{%
{Romero-Wolf}%
\ \protect \BOthers {.}}{%
{\protect \APACyear {2015}}%
}]{%
Romero-Wolf_2015}
\APACinsertmetastar {%
Romero-Wolf_2015}%
\begin{APACrefauthors}%
{Romero-Wolf}, A.%
, {Vance}, S.%
, {Maiwald}, F.%
, {Heggy}, E.%
, {Ries}, P.%
\BCBL {}\ \BBA {} {Liewer}, K.%
\end{APACrefauthors}%
\unskip\
\newblock
\APACrefYearMonthDay{2015}{{\APACmonth{03}}}{}.
\newblock
{\BBOQ}\APACrefatitle {{A passive probe for subsurface oceans and liquid water
  in Jupiter's icy moons}} {{A passive probe for subsurface oceans and liquid
  water in Jupiter's icy moons}}.{\BBCQ}
\newblock
\APACjournalVolNumPages{Icarus}{248}{}{463-477}.
\newblock
\begin{APACrefDOI} \doi{10.1016/j.icarus.2014.10.04310.48550/arXiv.1404.1876}
  \end{APACrefDOI}
\PrintBackRefs{\CurrentBib}

\bibitem [\protect \citeauthoryear {%
Schenk%
\ \BBA {} Moore%
}{%
Schenk%
\ \BBA {} Moore%
}{%
{\protect \APACyear {2020}}%
}]{%
schenk2020}
\APACinsertmetastar {%
schenk2020}%
\begin{APACrefauthors}%
Schenk, P\BPBI M.%
\BCBT {}\ \BBA {} Moore, J\BPBI M.%
\end{APACrefauthors}%
\unskip\
\newblock
\APACrefYearMonthDay{2020}{}{}.
\newblock
{\BBOQ}\APACrefatitle {Topography and geology of Uranian mid-sized icy
  satellites in comparison with Saturnian and Plutonian satellites} {Topography
  and geology of uranian mid-sized icy satellites in comparison with saturnian
  and plutonian satellites}.{\BBCQ}
\newblock
\APACjournalVolNumPages{Philosophical Transactions of the Royal Society
  A}{378}{2187}{20200102}.
\PrintBackRefs{\CurrentBib}

\bibitem [\protect \citeauthoryear {%
{Schroeder}%
\ \protect \BOthers {.}}{%
{Schroeder}%
\ \protect \BOthers {.}}{%
{\protect \APACyear {2016}}%
}]{%
Schroeder_2016}
\APACinsertmetastar {%
Schroeder_2016}%
\begin{APACrefauthors}%
{Schroeder}, D\BPBI M.%
, {Romero-Wolf}, A.%
, {Carrer}, L.%
, {Grima}, C.%
, {Campbell}, B\BPBI A.%
, {Kofman}, W.%
\BDBL {}{Blankenship}, D\BPBI D.%
\end{APACrefauthors}%
\unskip\
\newblock
\APACrefYearMonthDay{2016}{{\APACmonth{12}}}{}.
\newblock
{\BBOQ}\APACrefatitle {{Assessing the potential for passive radio sounding of
  Europa and Ganymede with RIME and REASON}} {{Assessing the potential for
  passive radio sounding of Europa and Ganymede with RIME and REASON}}.{\BBCQ}
\newblock
\APACjournalVolNumPages{Planetary and Space Science}{134}{}{52-60}.
\newblock
\begin{APACrefDOI} \doi{10.1016/j.pss.2016.10.007} \end{APACrefDOI}
\PrintBackRefs{\CurrentBib}

\bibitem [\protect \citeauthoryear {%
{Sittler}%
, {Ogilvie}%
\BCBL {}\ \BBA {} {Selesnick}%
}{%
{Sittler}%
\ \protect \BOthers {.}}{%
{\protect \APACyear {1987}}%
}]{%
Sittler_1987}
\APACinsertmetastar {%
Sittler_1987}%
\begin{APACrefauthors}%
{Sittler}, J., Edward~C.%
, {Ogilvie}, K\BPBI W.%
\BCBL {}\ \BBA {} {Selesnick}, R.%
\end{APACrefauthors}%
\unskip\
\newblock
\APACrefYearMonthDay{1987}{{\APACmonth{12}}}{}.
\newblock
{\BBOQ}\APACrefatitle {{Survey of electrons in the Uranian magnetosphere:
  Voyager 2 observations}} {{Survey of electrons in the Uranian magnetosphere:
  Voyager 2 observations}}.{\BBCQ}
\newblock
\APACjournalVolNumPages{Journal of Geophysical Research}{92}{A13}{15263-15281}.
\newblock
\begin{APACrefDOI} \doi{10.1029/JA092iA13p15263} \end{APACrefDOI}
\PrintBackRefs{\CurrentBib}

\bibitem [\protect \citeauthoryear {%
Souček%
\ \protect \BOthers {.}}{%
Souček%
\ \protect \BOthers {.}}{%
{\protect \APACyear {2023}}%
}]{%
Soucek2023}
\APACinsertmetastar {%
Soucek2023}%
\begin{APACrefauthors}%
Souček, O.%
, Běhounková, M.%
, Schroeder, D\BPBI M.%
, Wolfenbarger, N\BPBI S.%
, Kalousová, K.%
, Steinbrügge, G.%
\BCBL {}\ \BBA {} Soderlund, K\BPBI M.%
\end{APACrefauthors}%
\unskip\
\newblock
\APACrefYearMonthDay{2023}{}{}.
\newblock
{\BBOQ}\APACrefatitle {Radar Attenuation in Enceladus’ Ice Shell: Obstacles
  and Opportunities for Constraining Shell Thickness, Chemistry, and Thermal
  Structure} {Radar attenuation in enceladus’ ice shell: Obstacles and
  opportunities for constraining shell thickness, chemistry, and thermal
  structure}.{\BBCQ}
\newblock
\APACjournalVolNumPages{Journal of Geophysical Research:
  Planets}{}{}{e2022JE007626}.
\newblock
\begin{APACrefDOI} \doi{https://doi.org/10.1029/2022JE007626} \end{APACrefDOI}
\PrintBackRefs{\CurrentBib}

\bibitem [\protect \citeauthoryear {%
Steinbrügge%
\ \protect \BOthers {.}}{%
Steinbrügge%
\ \protect \BOthers {.}}{%
{\protect \APACyear {2021}}%
}]{%
Steinbruegge_2021}
\APACinsertmetastar {%
Steinbruegge_2021}%
\begin{APACrefauthors}%
Steinbrügge, G.%
, Romero-Wolf, A.%
, Peters, S.%
, Schroeder, D\BPBI M.%
, Carrer, L.%
, Hamilton, C\BPBI W.%
\BDBL {}Young, D\BPBI A.%
\end{APACrefauthors}%
\unskip\
\newblock
\APACrefYearMonthDay{2021}{mar 18}{}.
\newblock
{\BBOQ}\APACrefatitle {PRIME --- {A} {Passive} {Radar} {Sounding} {Concept} for
  {Io}} {Prime --- {A} {Passive} {Radar} {Sounding} {Concept} for {Io}}.{\BBCQ}
\newblock
\APACjournalVolNumPages{Bulletin of the AAS}{53}{4}{}.
\newblock
\APACrefnote{https://baas.aas.org/pub/2021n4i271}
\PrintBackRefs{\CurrentBib}

\bibitem [\protect \citeauthoryear {%
Stogryn%
\ \BBA {} Desargant%
}{%
Stogryn%
\ \BBA {} Desargant%
}{%
{\protect \APACyear {1985}}%
}]{%
stogryn1985}
\APACinsertmetastar {%
stogryn1985}%
\begin{APACrefauthors}%
Stogryn, A.%
\BCBT {}\ \BBA {} Desargant, G.%
\end{APACrefauthors}%
\unskip\
\newblock
\APACrefYearMonthDay{1985}{}{}.
\newblock
{\BBOQ}\APACrefatitle {The dielectric properties of brine in sea ice at
  microwave frequencies} {The dielectric properties of brine in sea ice at
  microwave frequencies}.{\BBCQ}
\newblock
\APACjournalVolNumPages{IEEE Transactions on Antennas and
  Propagation}{33}{5}{523--532}.
\PrintBackRefs{\CurrentBib}

\bibitem [\protect \citeauthoryear {%
{Stone}%
}{%
{Stone}%
}{%
{\protect \APACyear {1987}}%
}]{%
Stone_1987}
\APACinsertmetastar {%
Stone_1987}%
\begin{APACrefauthors}%
{Stone}, E\BPBI C.%
\end{APACrefauthors}%
\unskip\
\newblock
\APACrefYearMonthDay{1987}{{\APACmonth{12}}}{}.
\newblock
{\BBOQ}\APACrefatitle {{The Voyager 2 encounter with Uranus}} {{The Voyager 2
  encounter with Uranus}}.{\BBCQ}
\newblock
\APACjournalVolNumPages{Journal of Geophysical Research}{92}{A13}{14873-14876}.
\newblock
\begin{APACrefDOI} \doi{10.1029/JA092iA13p14873} \end{APACrefDOI}
\PrintBackRefs{\CurrentBib}

\bibitem [\protect \citeauthoryear {%
van~de Hulst%
}{%
van~de Hulst%
}{%
{\protect \APACyear {1981}}%
}]{%
hulst1981}
\APACinsertmetastar {%
hulst1981}%
\begin{APACrefauthors}%
van~de Hulst, H\BPBI C.%
\end{APACrefauthors}%
\unskip\
\newblock
\APACrefYear{1981}.
\newblock
\APACrefbtitle {Light scattering by small particles} {Light scattering by small
  particles}.
\newblock
\APACaddressPublisher{}{Courier Corporation}.
\PrintBackRefs{\CurrentBib}

\bibitem [\protect \citeauthoryear {%
{Weiss}%
\ \protect \BOthers {.}}{%
{Weiss}%
\ \protect \BOthers {.}}{%
{\protect \APACyear {2021}}%
}]{%
Weiss_2021}
\APACinsertmetastar {%
Weiss_2021}%
\begin{APACrefauthors}%
{Weiss}, B\BPBI P.%
, {Biersteker}, J\BPBI B.%
, {Colicci}, V.%
, {Goode}, A.%
, {Castillo-Rogez}, J\BPBI C.%
, {Petropoulos}, A\BPBI E.%
\BCBL {}\ \BBA {} {Balint}, T\BPBI S.%
\end{APACrefauthors}%
\unskip\
\newblock
\APACrefYearMonthDay{2021}{{\APACmonth{10}}}{}.
\newblock
{\BBOQ}\APACrefatitle {{Searching for Subsurface Oceans on the Moons of Uranus
  Using Magnetic Induction}} {{Searching for Subsurface Oceans on the Moons of
  Uranus Using Magnetic Induction}}.{\BBCQ}
\newblock
\APACjournalVolNumPages{Geophysical Research Letters}{48}{19}{e94758}.
\newblock
\begin{APACrefDOI} \doi{10.1029/2021GL094758} \end{APACrefDOI}
\PrintBackRefs{\CurrentBib}

\bibitem [\protect \citeauthoryear {%
Wolfenbarger%
, Fox-Powell%
, Buffo%
, Soderlund%
\BCBL {}\ \BBA {} Blankenship%
}{%
Wolfenbarger%
\ \protect \BOthers {.}}{%
{\protect \APACyear {2022}}%
}]{%
Wolfenbarger2022}
\APACinsertmetastar {%
Wolfenbarger2022}%
\begin{APACrefauthors}%
Wolfenbarger, N\BPBI S.%
, Fox-Powell, M\BPBI G.%
, Buffo, J\BPBI J.%
, Soderlund, K\BPBI M.%
\BCBL {}\ \BBA {} Blankenship, D\BPBI D.%
\end{APACrefauthors}%
\unskip\
\newblock
\APACrefYearMonthDay{2022}{}{}.
\newblock
{\BBOQ}\APACrefatitle {Compositional Controls on the Distribution of Brine in
  Europa's Ice Shell} {Compositional controls on the distribution of brine in
  europa's ice shell}.{\BBCQ}
\newblock
\APACjournalVolNumPages{Journal of Geophysical Research:
  Planets}{127}{9}{e2022JE007305}.
\newblock
\begin{APACrefDOI} \doi{https://doi.org/10.1029/2022JE007305} \end{APACrefDOI}
\PrintBackRefs{\CurrentBib}

\bibitem [\protect \citeauthoryear {%
{Zarka}%
}{%
{Zarka}%
}{%
{\protect \APACyear {1998}}%
}]{%
Zarka_1998}
\APACinsertmetastar {%
Zarka_1998}%
\begin{APACrefauthors}%
{Zarka}, P.%
\end{APACrefauthors}%
\unskip\
\newblock
\APACrefYearMonthDay{1998}{{\APACmonth{09}}}{}.
\newblock
{\BBOQ}\APACrefatitle {{Auroral radio emissions at the outer planets:
  Observations and theories}} {{Auroral radio emissions at the outer planets:
  Observations and theories}}.{\BBCQ}
\newblock
\APACjournalVolNumPages{Journal of Geophysical Research}{103}{E9}{20159-20194}.
\newblock
\begin{APACrefDOI} \doi{10.1029/98JE01323} \end{APACrefDOI}
\PrintBackRefs{\CurrentBib}

\bibitem [\protect \citeauthoryear {%
{Zarka}%
}{%
{Zarka}%
}{%
{\protect \APACyear {2004}}%
}]{%
Zarka_2004}
\APACinsertmetastar {%
Zarka_2004}%
\begin{APACrefauthors}%
{Zarka}, P.%
\end{APACrefauthors}%
\unskip\
\newblock
\APACrefYearMonthDay{2004}{{\APACmonth{01}}}{}.
\newblock
{\BBOQ}\APACrefatitle {{Radio and plasma waves at the outer planets}} {{Radio
  and plasma waves at the outer planets}}.{\BBCQ}
\newblock
\APACjournalVolNumPages{Advances in Space Research}{33}{11}{2045-2060}.
\newblock
\begin{APACrefDOI} \doi{10.1016/j.asr.2003.07.055} \end{APACrefDOI}
\PrintBackRefs{\CurrentBib}

\bibitem [\protect \citeauthoryear {%
{Zarka}%
, {Cecconi}%
\BCBL {}\ \BBA {} {Kurth}%
}{%
{Zarka}%
\ \protect \BOthers {.}}{%
{\protect \APACyear {2004}}%
}]{%
Zarka_2004b}
\APACinsertmetastar {%
Zarka_2004b}%
\begin{APACrefauthors}%
{Zarka}, P.%
, {Cecconi}, B.%
\BCBL {}\ \BBA {} {Kurth}, W\BPBI S.%
\end{APACrefauthors}%
\unskip\
\newblock
\APACrefYearMonthDay{2004}{{\APACmonth{09}}}{}.
\newblock
{\BBOQ}\APACrefatitle {{Jupiter's low-frequency radio spectrum from
  Cassini/Radio and Plasma Wave Science (RPWS) absolute flux density
  measurements}} {{Jupiter's low-frequency radio spectrum from Cassini/Radio
  and Plasma Wave Science (RPWS) absolute flux density measurements}}.{\BBCQ}
\newblock
\APACjournalVolNumPages{Journal of Geophysical Research (Space
  Physics)}{109}{A9}{A09S15}.
\newblock
\begin{APACrefDOI} \doi{10.1029/2003JA010260} \end{APACrefDOI}
\PrintBackRefs{\CurrentBib}

\end{thebibliography}
%




%
%
%
%
%

\end{document}